\documentclass[longauth]{aa} 

\usepackage{overpic}
\usepackage{booktabs}
\usepackage{tabularx}
\usepackage{multirow}
\usepackage{graphicx}
\usepackage{amssymb, amsmath}
\usepackage{color}
\usepackage{xspace}
\usepackage{siunitx}
\usepackage{url,hyperref}
\usepackage{footmisc}
\usepackage{subcaption}
\usepackage[switch]{lineno}

\usepackage[export]{adjustbox}

\newcommand{\ra}[1]{\renewcommand{\arraystretch}{#1}}
\newcommand{\psr}{PSR B1259--63\xspace}
\newcommand{\psrls}{\psr/LS 2883\xspace}
\newcommand{\hess}{H.E.S.S.\xspace}
\newcommand{\fl}{{\it Fermi}-LAT\xspace}
\newcommand{\vhe}{VHE\xspace}
\newcommand{\vhes}{VHEs\xspace}

\DeclareSIUnit\year{yr}
\DeclareSIUnit\erg{erg}
\DeclareSIUnit\au{AU}
\DeclareSIUnit{\msun}{\mbox{$\textrm{M}_{\odot}$}}
\DeclareSIUnit{\rsun}{\mbox{$R_{\odot}$}}
\DeclareSIUnit\ev{eV}
\DeclareSIUnit\kev{\kilo\ev}
\DeclareSIUnit\mev{\mega\ev}
\DeclareSIUnit\gev{\giga\ev}
\DeclareSIUnit\tev{\tera\ev}
\DeclareSIUnit\pe{p.e.}
\DeclareSIUnit\parsec{pc}
\DeclareSIUnit\gauss{G}
\DeclareSIUnit\lightyear{ly}
\DeclareSIUnit\day{d}
\DeclareSIUnit\photon{ph}

\def \hcm {\hbox {\ifmmode $ atom cm$^{-2}\else atom cm$^{-2}$\fi}}

\def \apjl {ApJL}
\def \apjs {ApJS}
\def \aap {A\&A}

\makeatletter
\renewcommand*\aa@pageof{, page \thepage{} of \pageref*{LastPage}}
\makeatother

\makeatletter
\renewcommand*{\@fnsymbol}[1]{\ifcase#1\or*\or$\dagger$\or$\ddagger$\or**\or$\dagger\dagger$\or$\ddagger\ddagger$\fi}
\makeatother

\begin{document}
\title{\hess and \fl observations of \psr/ \\LS 2883 during its 2014 and 2017 periastron passages}
\titlerunning{\psr in 2014 and 2017}
\authorrunning{H.E.S.S. Collaboration}
\date{}
\author{\small H.E.S.S. Collaboration
\and H.~Abdalla \inst{\ref{NWU}}
\and R.~Adam \inst{\ref{LLR}}
\and F.~Aharonian \inst{\ref{MPIK},\ref{DIAS},\ref{RAU}}
\and F.~Ait~Benkhali \inst{\ref{MPIK}}
\and E.O.~Ang\"uner \inst{\ref{CPPM}}
\and M.~Arakawa \inst{\ref{Rikkyo}}
\and C.~Arcaro \inst{\ref{NWU}}
\and C.~Armand \inst{\ref{LAPP}}
\and H.~Ashkar \inst{\ref{IRFU}}
\and M.~Backes \inst{\ref{UNAM},\ref{NWU}}
\and V.~Barbosa~Martins \inst{\ref{DESY}}
\and M.~Barnard \inst{\ref{NWU}}
\and Y.~Becherini \inst{\ref{Linnaeus}}
\and D.~Berge \inst{\ref{DESY}}
\and K.~Bernl\"ohr \inst{\ref{MPIK}}
\and R.~Blackwell \inst{\ref{Adelaide}}
\and M.~B\"ottcher \inst{\ref{NWU}}
\and C.~Boisson \inst{\ref{LUTH}}
\and J.~Bolmont \inst{\ref{LPNHE}}
\and S.~Bonnefoy \inst{\ref{DESY}}
\and J.~Bregeon \inst{\ref{LUPM}}
\and M.~Breuhaus \inst{\ref{MPIK}}
\and F.~Brun \inst{\ref{IRFU}}
\and P.~Brun \inst{\ref{IRFU}}
\and M.~Bryan \inst{\ref{GRAPPA}}
\and M.~B\"{u}chele \inst{\ref{ECAP}}
\and T.~Bulik \inst{\ref{UWarsaw}}
\and T.~Bylund \inst{\ref{Linnaeus}}
\and S.~Caroff \inst{\ref{LPNHE}}
\and A.~Carosi \inst{\ref{LAPP}}
\and S.~Casanova \inst{\ref{IFJPAN},\ref{MPIK}}
\and M.~Cerruti \inst{\ref{LPNHE},\ref{CerrutiNowAt}}
\and T.~Chand \inst{\ref{NWU}}
\and S.~Chandra \inst{\ref{NWU}}
\and R.C.G.~Chaves \inst{\ref{LUPM}}
\and A.~Chen \inst{\ref{WITS}}
\and S.~Colafrancesco \inst{\ref{WITS}} \protect\footnotemark[2]
\and M.~Cury{\l}o \inst{\ref{UWarsaw}}
\and I.D.~Davids \inst{\ref{UNAM}}
\and C.~Deil \inst{\ref{MPIK}}
\and J.~Devin \inst{\ref{CENBG}}
\and P.~deWilt \inst{\ref{Adelaide}}
\and L.~Dirson \inst{\ref{HH}}
\and A.~Djannati-Ata\"i \inst{\ref{APC}}
\and A.~Dmytriiev \inst{\ref{LUTH}}
\and A.~Donath \inst{\ref{MPIK}}
\and V.~Doroshenko \inst{\ref{IAAT}}
\and J.~Dyks \inst{\ref{NCAC}}
\and K.~Egberts \inst{\ref{UP}}
\and G.~Emery \inst{\ref{LPNHE}}
\and J.-P.~Ernenwein \inst{\ref{CPPM}}
\and S.~Eschbach \inst{\ref{ECAP}}
\and K.~Feijen \inst{\ref{Adelaide}}
\and S.~Fegan \inst{\ref{LLR}}
\and A.~Fiasson \inst{\ref{LAPP}}
\and G.~Fontaine \inst{\ref{LLR}}
\and S.~Funk \inst{\ref{ECAP}}
\and M.~F\"u{\ss}ling \inst{\ref{DESY}}
\and S.~Gabici \inst{\ref{APC}}
\and Y.A.~Gallant \inst{\ref{LUPM}}
\and F.~Gat{\'e} \inst{\ref{LAPP}}
\and G.~Giavitto \inst{\ref{DESY}}
\and L.~Giunti \inst{\ref{APC}}
\and D.~Glawion \inst{\ref{LSW}}
\and J.F.~Glicenstein \inst{\ref{IRFU}}
\and D.~Gottschall \inst{\ref{IAAT}}
\and M.-H.~Grondin \inst{\ref{CENBG}}
\and J.~Hahn \inst{\ref{MPIK}}
\and M.~Haupt \inst{\ref{DESY}}
\and G.~Heinzelmann \inst{\ref{HH}}
\and G.~Henri \inst{\ref{Grenoble}}
\and G.~Hermann \inst{\ref{MPIK}}
\and J.A.~Hinton \inst{\ref{MPIK}}
\and W.~Hofmann \inst{\ref{MPIK}}
\and C.~Hoischen \inst{\ref{UP}}
\and T.~L.~Holch \inst{\ref{HUB}}
\and M.~Holler \inst{\ref{LFUI}}
\and D.~Horns \inst{\ref{HH}}
\and D.~Huber \inst{\ref{LFUI}}
\and H.~Iwasaki \inst{\ref{Rikkyo}}
\and M.~Jamrozy \inst{\ref{UJK}}
\and D.~Jankowsky \inst{\ref{ECAP}}
\and F.~Jankowsky \inst{\ref{LSW}}
\and A.~Jardin-Blicq \inst{\ref{MPIK}}
\and I.~Jung-Richardt \inst{\ref{ECAP}}
\and M.A.~Kastendieck \inst{\ref{HH}}
\and K.~Katarzy{\'n}ski \inst{\ref{NCUT}}
\and M.~Katsuragawa \inst{\ref{KAVLI}}
\and U.~Katz \inst{\ref{ECAP}}
\and D.~Khangulyan \inst{\ref{Rikkyo}}
\and B.~Kh\'elifi \inst{\ref{APC}}
\and J.~King \inst{\ref{LSW}}
\and S.~Klepser \inst{\ref{DESY}}
\and W.~Klu\'{z}niak \inst{\ref{NCAC}}
\and Nu.~Komin \inst{\ref{WITS}}
\and K.~Kosack \inst{\ref{IRFU}}
\and D.~Kostunin \inst{\ref{DESY}}
\and M.~Kreter \inst{\ref{NWU}}
\and G.~Lamanna \inst{\ref{LAPP}}
\and A.~Lemi\`ere \inst{\ref{APC}}
\and M.~Lemoine-Goumard \inst{\ref{CENBG}}
\and J.-P.~Lenain \inst{\ref{LPNHE}}
\and E.~Leser \inst{\ref{UP},\ref{DESY}}
\and C.~Levy \inst{\ref{LPNHE}}
\and T.~Lohse \inst{\ref{HUB}}
\and I.~Lypova \inst{\ref{DESY}}
\and J.~Mackey \inst{\ref{DIAS}}
\and J.~Majumdar \inst{\ref{DESY}}
\and D.~Malyshev \inst{\ref{IAAT}}
\and D.~Malyshev \inst{\ref{ECAP}}
\and V.~Marandon \inst{\ref{MPIK}}
\and A.~Marcowith \inst{\ref{LUPM}}
\and A.~Mares \inst{\ref{CENBG}}
\and C.~Mariaud* \inst{\ref{LLR}}
\and G.~Mart\'i-Devesa \inst{\ref{LFUI}}
\and R.~Marx \inst{\ref{MPIK}}
\and G.~Maurin \inst{\ref{LAPP}}
\and P.J.~Meintjes \inst{\ref{UFS}}
\and A.M.W.~Mitchell \inst{\ref{MPIK},\ref{MitchellNowAt}}
\and R.~Moderski \inst{\ref{NCAC}}
\and M.~Mohamed \inst{\ref{LSW}}
\and L.~Mohrmann \inst{\ref{ECAP}}
\and C.~Moore \inst{\ref{Leicester}}
\and E.~Moulin \inst{\ref{IRFU}}
\and J.~Muller \inst{\ref{LLR}}
\and T.~Murach* \inst{\ref{DESY}}
\and S.~Nakashima  \inst{\ref{RIKKEN}}
\and M.~de~Naurois \inst{\ref{LLR}}
\and H.~Ndiyavala  \inst{\ref{NWU}}
\and F.~Niederwanger \inst{\ref{LFUI}}
\and J.~Niemiec \inst{\ref{IFJPAN}}
\and L.~Oakes \inst{\ref{HUB}}
\and P.~O'Brien \inst{\ref{Leicester}}
\and H.~Odaka \inst{\ref{Tokyo}}
\and S.~Ohm \inst{\ref{DESY}}
\and E.~de~Ona~Wilhelmi \inst{\ref{DESY}}
\and M.~Ostrowski \inst{\ref{UJK}}
\and I.~Oya \inst{\ref{DESY}}
\and M.~Panter \inst{\ref{MPIK}}
\and R.D.~Parsons \inst{\ref{MPIK}}
\and C.~Perennes \inst{\ref{LPNHE}}
\and P.-O.~Petrucci \inst{\ref{Grenoble}}
\and B.~Peyaud \inst{\ref{IRFU}}
\and Q.~Piel \inst{\ref{LAPP}}
\and S.~Pita \inst{\ref{APC}}
\and V.~Poireau \inst{\ref{LAPP}}
\and A.~Priyana~Noel \inst{\ref{UJK}}
\and D.A.~Prokhorov \inst{\ref{WITS}}
\and H.~Prokoph \inst{\ref{DESY}}
\and G.~P\"uhlhofer \inst{\ref{IAAT}}
\and M.~Punch \inst{\ref{APC},\ref{Linnaeus}}
\and A.~Quirrenbach \inst{\ref{LSW}}
\and S.~Raab \inst{\ref{ECAP}}
\and R.~Rauth \inst{\ref{LFUI}}
\and A.~Reimer \inst{\ref{LFUI}}
\and O.~Reimer \inst{\ref{LFUI}}
\and Q.~Remy \inst{\ref{LUPM}}
\and M.~Renaud \inst{\ref{LUPM}}
\and F.~Rieger \inst{\ref{MPIK}}
\and L.~Rinchiuso \inst{\ref{IRFU}}
\and C.~Romoli* \inst{\ref{MPIK}}
\and G.~Rowell \inst{\ref{Adelaide}}
\and B.~Rudak \inst{\ref{NCAC}}
\and E.~Ruiz-Velasco \inst{\ref{MPIK}}
\and V.~Sahakian \inst{\ref{YPI}}
\and S.~Sailer \inst{\ref{MPIK}}
\and S.~Saito \inst{\ref{Rikkyo}}
\and D.A.~Sanchez \inst{\ref{LAPP}}
\and A.~Santangelo \inst{\ref{IAAT}}
\and M.~Sasaki \inst{\ref{ECAP}}
\and R.~Schlickeiser \inst{\ref{RUB}}
\and F.~Sch\"ussler \inst{\ref{IRFU}}
\and A.~Schulz \inst{\ref{DESY}}
\and H.M.~Schutte \inst{\ref{NWU}}
\and U.~Schwanke \inst{\ref{HUB}}
\and S.~Schwemmer \inst{\ref{LSW}}
\and M.~Seglar-Arroyo \inst{\ref{IRFU}}
\and M.~Senniappan \inst{\ref{Linnaeus}}
\and A.S.~Seyffert \inst{\ref{NWU}}
\and N.~Shafi \inst{\ref{WITS}}
\and K.~Shiningayamwe \inst{\ref{UNAM}}
\and R.~Simoni \inst{\ref{GRAPPA}}
\and A.~Sinha \inst{\ref{APC}}
\and H.~Sol \inst{\ref{LUTH}}
\and A.~Specovius \inst{\ref{ECAP}}
\and M.~Spir-Jacob \inst{\ref{APC}}
\and {\L.}~Stawarz \inst{\ref{UJK}}
\and R.~Steenkamp \inst{\ref{UNAM}}
\and C.~Stegmann \inst{\ref{UP},\ref{DESY}}
\and C.~Steppa \inst{\ref{UP}}
\and T.~Takahashi  \inst{\ref{KAVLI}}
\and T.~Tavernier \inst{\ref{IRFU}}
\and A.M.~Taylor \inst{\ref{DESY}}
\and R.~Terrier \inst{\ref{APC}}
\and D.~Tiziani \inst{\ref{ECAP}}
\and M.~Tluczykont \inst{\ref{HH}}
\and C.~Trichard \inst{\ref{LLR}}
\and M.~Tsirou \inst{\ref{LUPM}}
\and N.~Tsuji \inst{\ref{Rikkyo}}
\and R.~Tuffs \inst{\ref{MPIK}}
\and Y.~Uchiyama \inst{\ref{Rikkyo}}
\and D.J.~van~der~Walt \inst{\ref{NWU}}
\and C.~van~Eldik \inst{\ref{ECAP}}
\and C.~van~Rensburg \inst{\ref{NWU}}
\and B.~van~Soelen \inst{\ref{UFS}}
\and G.~Vasileiadis \inst{\ref{LUPM}}
\and J.~Veh \inst{\ref{ECAP}}
\and C.~Venter \inst{\ref{NWU}}
\and P.~Vincent \inst{\ref{LPNHE}}
\and J.~Vink \inst{\ref{GRAPPA}}
\and H.J.~V\"olk \inst{\ref{MPIK}}
\and T.~Vuillaume \inst{\ref{LAPP}}
\and Z.~Wadiasingh \inst{\ref{NWU}}
\and S.J.~Wagner \inst{\ref{LSW}}
\and R.~White \inst{\ref{MPIK}}
\and A.~Wierzcholska \inst{\ref{IFJPAN},\ref{LSW}}
\and R.~Yang \inst{\ref{MPIK}}
\and H.~Yoneda \inst{\ref{KAVLI}}
\and M.~Zacharias \inst{\ref{NWU}}
\and R.~Zanin \inst{\ref{MPIK}}
\and A.A.~Zdziarski \inst{\ref{NCAC}}
\and A.~Zech \inst{\ref{LUTH}}
\and J.~Zorn \inst{\ref{MPIK}}
\and N.~\.Zywucka \inst{\ref{NWU}}
\\\vspace{\lineskip}and P.~Bordas* \inst{\ref{CerrutiNowAt}}
}

\institute{
Centre for Space Research, North-West University, Potchefstroom 2520, South Africa \label{NWU} \and
Universit\"at Hamburg, Institut f\"ur Experimentalphysik, Luruper Chaussee 149, D 22761 Hamburg, Germany \label{HH} \and
Max-Planck-Institut f\"ur Kernphysik, P.O. Box 103980, D 69029 Heidelberg, Germany \label{MPIK} \and
Dublin Institute for Advanced Studies, 31 Fitzwilliam Place, Dublin 2, Ireland \label{DIAS} \and
High Energy Astrophysics Laboratory, RAU,  123 Hovsep Emin St  Yerevan 0051, Armenia \label{RAU} \and
Yerevan Physics Institute, 2 Alikhanian Brothers St., 375036 Yerevan, Armenia \label{YPI} \and
Institut f\"ur Physik, Humboldt-Universit\"at zu Berlin, Newtonstr. 15, D 12489 Berlin, Germany \label{HUB} \and
University of Namibia, Department of Physics, Private Bag 13301, Windhoek, Namibia, 12010 \label{UNAM} \and
GRAPPA, Anton Pannekoek Institute for Astronomy, University of Amsterdam,  Science Park 904, 1098 XH Amsterdam, The Netherlands \label{GRAPPA} \and
Department of Physics and Electrical Engineering, Linnaeus University,  351 95 V\"axj\"o, Sweden \label{Linnaeus} \and
Institut f\"ur Theoretische Physik, Lehrstuhl IV: Weltraum und Astrophysik, Ruhr-Universit\"at Bochum, D 44780 Bochum, Germany \label{RUB} \and
Institut f\"ur Astro- und Teilchenphysik, Leopold-Franzens-Universit\"at Innsbruck, A-6020 Innsbruck, Austria \label{LFUI} \and
School of Physical Sciences, University of Adelaide, Adelaide 5005, Australia \label{Adelaide} \and
LUTH, Observatoire de Paris, PSL Research University, CNRS, Universit\'e Paris Diderot, 5 Place Jules Janssen, 92190 Meudon, France \label{LUTH} \and
Sorbonne Universit\'e, Universit\'e Paris Diderot, Sorbonne Paris Cit\'e, CNRS/IN2P3, Laboratoire de Physique Nucl\'eaire et de Hautes Energies, LPNHE, 4 Place Jussieu, F-75252 Paris, France \label{LPNHE} \and
Laboratoire Univers et Particules de Montpellier, Universit\'e Montpellier, CNRS/IN2P3,  CC 72, Place Eug\`ene Bataillon, F-34095 Montpellier Cedex 5, France \label{LUPM} \and
IRFU, CEA, Universit\'e Paris-Saclay, F-91191 Gif-sur-Yvette, France \label{IRFU} \and
Astronomical Observatory, The University of Warsaw, Al. Ujazdowskie 4, 00-478 Warsaw, Poland \label{UWarsaw} \and
Aix Marseille Universit\'e, CNRS/IN2P3, CPPM, Marseille, France \label{CPPM} \and
Instytut Fizyki J\c{a}drowej PAN, ul. Radzikowskiego 152, 31-342 Krak{\'o}w, Poland \label{IFJPAN} \and
School of Physics, University of the Witwatersrand, 1 Jan Smuts Avenue, Braamfontein, Johannesburg, 2050 South Africa \label{WITS} \and
Laboratoire d'Annecy de Physique des Particules, Univ. Grenoble Alpes, Univ. Savoie Mont Blanc, CNRS, LAPP, 74000 Annecy, France \label{LAPP} \and
Landessternwarte, Universit\"at Heidelberg, K\"onigstuhl, D 69117 Heidelberg, Germany \label{LSW} \and
Universit\'e Bordeaux, CNRS/IN2P3, Centre d'\'Etudes Nucl\'eaires de Bordeaux Gradignan, 33175 Gradignan, France \label{CENBG} \and
Institut f\"ur Astronomie und Astrophysik, Universit\"at T\"ubingen, Sand 1, D 72076 T\"ubingen, Germany \label{IAAT} \and
Laboratoire Leprince-Ringuet, École Polytechnique, CNRS, Institut Polytechnique de Paris, F-91128 Palaiseau, France \label{LLR} \and
APC, AstroParticule et Cosmologie, Universit\'{e} Paris Diderot, CNRS/IN2P3, CEA/Irfu, Observatoire de Paris, Sorbonne Paris Cit\'{e}, 10, rue Alice Domon et L\'{e}onie Duquet, 75205 Paris Cedex 13, France \label{APC} \and
Univ. Grenoble Alpes, CNRS, IPAG, F-38000 Grenoble, France \label{Grenoble} \and
Department of Physics and Astronomy, The University of Leicester, University Road, Leicester, LE1 7RH, United Kingdom \label{Leicester} \and
Nicolaus Copernicus Astronomical Center, Polish Academy of Sciences, ul. Bartycka 18, 00-716 Warsaw, Poland \label{NCAC} \and
Institut f\"ur Physik und Astronomie, Universit\"at Potsdam,  Karl-Liebknecht-Strasse 24/25, D 14476 Potsdam, Germany \label{UP} \and
Friedrich-Alexander-Universit\"at Erlangen-N\"urnberg, Erlangen Centre for Astroparticle Physics, Erwin-Rommel-Str. 1, D 91058 Erlangen, Germany \label{ECAP} \and
DESY, D-15738 Zeuthen, Germany \label{DESY} \and
Obserwatorium Astronomiczne, Uniwersytet Jagiello{\'n}ski, ul. Orla 171, 30-244 Krak{\'o}w, Poland \label{UJK} \and
Centre for Astronomy, Faculty of Physics, Astronomy and Informatics, Nicolaus Copernicus University,  Grudziadzka 5, 87-100 Torun, Poland \label{NCUT} \and
Department of Physics, University of the Free State,  PO Box 339, Bloemfontein 9300, South Africa \label{UFS} \and
Department of Physics, Rikkyo University, 3-34-1 Nishi-Ikebukuro, Toshima-ku, Tokyo 171-8501, Japan \label{Rikkyo} \and
Kavli Institute for the Physics and Mathematics of the Universe (WPI), The University of Tokyo Institutes for Advanced Study (UTIAS), The University of Tokyo, 5-1-5 Kashiwa-no-Ha, Kashiwa, Chiba, 277-8583, Japan \label{KAVLI} \and
Department of Physics, The University of Tokyo, 7-3-1 Hongo, Bunkyo-ku, Tokyo 113-0033, Japan \label{Tokyo} \and
RIKEN, 2-1 Hirosawa, Wako, Saitama 351-0198, Japan \label{RIKKEN} \and
Now at Physik Institut, Universit\"at Z\"urich, Winterthurerstrasse 190, CH-8057 Z\"urich, Switzerland \label{MitchellNowAt} \and
Now at Institut de Ci\`{e}ncies del Cosmos (ICC UB), Universitat de Barcelona (IEEC-UB), Mart\'{i} Franqu\`es 1, E08028 Barcelona, Spain \label{CerrutiNowAt}
}

\offprints{H.E.S.S.~collaboration,
\protect\\\email{\href{mailto:contact.hess@hess-experiment.eu}{contact.hess@hess-experiment.eu}};
\protect\\\protect\footnotemark[1] Corresponding authors
\protect\\\protect\footnotemark[2] Deceased
}


\abstract{
\psrls is a gamma-ray binary system consisting of a pulsar in an eccentric orbit around a
bright O{\it{e}} stellar-type companion star that features a dense circumstellar disc. The bright
broad-band emission observed at phases close to periastron offers a
unique opportunity to study particle acceleration and radiation processes in binary systems.
Observations at gamma-ray energies constrain these
processes through variability and spectral characterisation studies.}
%
%
{The high- and very-high-energy (HE, VHE) gamma-ray emission from \psrls around the times of its periastron passage are characterised,
in particular, at the time of the HE gamma-ray flares reported to have occurred in 2011, 2014, and 2017.
Short-term and average emission characteristics of \psrls are determined. Super-orbital variability is searched for in order to investigate
possible cycle-to-cycle VHE flux changes due to different properties of the companion star's circumstellar disc and/or the conditions
under which the HE gamma-ray flares develop.}
%
%
{Spectra and light curves were derived from observations conducted with the \hess-II array in 2014 and 2017.
Phase-folded light curves are compared with the results obtained
in 2004, 2007, and 2011. \fl observations from 2010/11, 2014, and 2017 are analysed.}
%
%
{A local double-peak profile with asymmetric peaks in the VHE light curve is measured, with
a flux minimum at the time of periastron $t_p$ and two peaks coinciding
with the times at which the neutron star crosses the companion's circumstellar disc ($\sim t_p \pm \SI{16}{\day}$). A high VHE
gamma-ray flux is also observed at the times of the HE gamma-ray flares ($\sim t_{\rm p}
+ \SI{30}{\day}$) and at phases before the first disc crossing ($\sim t_{\rm p} - \SI{35}{\day}$).
The spectral energy range now extends to below \SI{200}{\gev} and up to $\sim \SI{45}{\tev}$.}
%
%
{\psrls displays periodic flux variability at VHE gamma-rays without clear signatures of super-orbital modulation in the time span covered by the monitoring of the source with the H.E.S.S. telescopes. This flux variability is most probably caused by the changing environmental conditions, particularly at times close to periastron passage at which the neutron star is thought to cross the circumstellar disc of the companion star twice. In contrast, the photon index remains unchanged within uncertainties for about \SI{200}{\day} around periastron. At HE gamma-rays, \psrls has now been detected also before and after periastron, close to the disc crossing times. Repetitive flares with distinct variability patterns are detected in this energy range. Such outbursts are not observed at VHEs, although a relatively high emission level is measured. The spectra obtained in both energy regimes displays a similar slope, although a common physical origin either in terms of a related particle population, emission mechanism, or emitter location is ruled out.}




\maketitle

\makeatletter
\renewcommand*{\@fnsymbol}[1]{\ifcase#1\@arabic{#1}\fi}
\makeatother



\section{Introduction}
\label{section:introduction}


Gamma-ray binaries consist of a massive star and a compact object, either a stellar-mass black
hole or a neutron star, in orbit around each other. These systems display a non-thermal energy flux maximum
in the gamma-ray band. At very high energies (VHE; $E>\SI{100}{\gev}$) only a handful of such systems have been detected:
LS 5039 \citep{Aharonian2005a}, \psrls \citep{Aharonian2005b}, LS I +61 303 \citep{Albert2006},
HESS J0632+057 \citep{Aharonian2007}, HESS J1018-589 \citep{HESS2015}, and the recently discovered objects LMC P3
\citep{LMC_P3} and TeV J2032+4130 \citep{MAGIC_VERITAS2017}. Only in the cases of \psrls and TeV J2032+4130 are the compact
companions well-identified, in both cases as pulsars, making them unique objects for the study of the
interaction between pulsar and stellar winds as well as particle acceleration as well as emission and absorption
mechanisms in close binary systems.

Initially discovered in a high-frequency radio survey aiming to detect young and distant short-period
pulsars, \psrls has since been the object of extensive studies at all frequencies. The source is
composed of the rapidly rotating pulsar \psrls with a spin period of $\SI{48}{\milli\second}$ and a spin-down luminosity
of $\SI{8e35}{\erg\per\second}$ and a bright O{\it{e}} companion star, LS 2883, with a
bolometric luminosity of $L_{\ast}=\SI{2.3e38}{\erg\per\second}$ \citep{Negueruela2011}.

The pulsar orbits the companion with a period $P_{\mathrm{orb}} = \SI{3.4}{\year}$ (\SI{1237}{\day}) in a very eccentric orbit ($e = 0.87$) with an orbital
separation of about 13.4 astronomical units (\si{\au}s) at apastron and less than \SI{1}{\au} at periastron
\citep{Johnston1992b,Wex1998,Wang2004}. The mass function of the system indicates a mass of the companion star
of $M_{\ast} \approx \SI{30}{\msun}$ and an orbital inclination angle $i_{\mathrm{orb}} \approx\ang{25}$ for
the minimal neutron star mass of $\SI{1.4}{\msun}$.

The massive star LS 2883
features an equatorial disc which extends to at least 10 stellar radii \citep{Johnston1992b,Negueruela2011,Chernyakova2014}.
The disc is inclined with respect to the pulsar's orbital plane \citep{Johnston1992b, melatos1995, Negueruela2011}
in such a way that the pulsar crosses the disc twice each orbit, just before ($\sim \SI{16}{\day}$) and after
($\sim \SI{16}{\day}$, \citealp{Johnston2005}) the time of periastron ($t_{\rm p}$).

\psrls displays broad-band emission which extends from radio wavelengths up to VHE gamma rays. In the radio
domain, a pulsed component is detectable until the system approaches periastron (see e.g. \citealp{Johnston1992}).
Thereafter the intensity of the radio pulsed emission decreases until its complete disappearance between $t_{\rm p}-\SI{16}{\day}$
and $t_{\rm p}+\SI{16}{\day}$. The definition of the nominal times of the disc crossings is based on the
disappearance of this pulse. The times of the disc crossings are almost constant across orbits, although some differences of a few days
have been reported (see e.g. \citealp{Connors2002, Johnston2005, Abdo2011}). A transient
unpulsed component appears and sharply rises to a level more than ten times higher than the flux density of the
pulsed emission far from periastron (see e.g. \citealp{Johnston2005}). Flux maxima are observed around the first and second disc crossings by the
neutron star. A similar behaviour is observed in the X-ray domain (see e.g. \citealp{Chernyakova2014}). The transient, unpulsed radio emission is presumably
synchrotron radiation produced by electrons accelerated at the shock interface between the relativistic pulsar
wind and stellar outflows (see e.g. \citep{Ball1999}).

At optical wavelengths, variability in the H$\alpha$ and HeI lines has been reported to occur around the periastron passage of \psrls, with the line strengths displaying a maximum about \SI{13}{\day} after periastron \citep{vanSoelen2016}. These observations are consistent with a scenario in which the circumstellar disc gets disrupted around periastron due to the interaction with the pulsar wind. Furthermore, the decrease in the equivalent width of the H$\alpha$ line seems to roughly coincide with the onset of the HE gamma-ray flare observed with the $Fermi$-LAT, implying a physical connection between the properties of the disc and the mechanisms behind the HE flares \citep{Chernyakova2015}.


In high-energy (HE; $\SI{0.1}{\gev} < E < \SI{100}{\gev}$) gamma rays, the source was detected by the
\textit{Fermi}-LAT for the first time around the
periastron passage of 2010/2011. In this energy band, a strong, unexpected enhancement of the emission
starting approximately \SI{30}{\day} after periastron was detected. This flare lasted more than one
month. At the peak, the emitted power in gamma rays almost matched the total spin-down luminosity of the pulsar
\citep{Abdo2011}. This flaring event was detected again around the periastron passages in 2014
(see e.g. \citealp{Caliandro2015}) and in 2017 (e.g. \citealp{Tam2018}), strengthening the hypothesis of a periodic phenomenon. The nature
of these flaring episodes is still unclear, with theoretical interpretations considering either the unshocked
pulsar wind, Doppler-boosted emission from shocked material, and/or enhanced photon field energy
densities provided by the circumstellar disc \citep{Chernyakova2015, Khangulyan2011, Khangulyan2012}.
Moreover, differences in the HE flare as observed in 2011, 2014 and in particular in 2017,
seem to indicate that other factors need to be accounted for to characterise it.
In particular, the flaring event of 2017 has shown variability characteristics such as minute scale variability
that were not identified in previous cycles \citep{Johnson2018}.

At very high energies, the \hess telescopes detected the source during the periastron passage in
2004 \citep{Aharonian2005b} and also recorded the subsequent passages in 2007 \citep{Aharonian2009},
2010/2011 \citep{HESS2013}, 2014 (see \cite{Romoli2015} for preliminary results of these observations)
and 2017. The source was detected firmly in every case except 2017, where the limited data set
prevented a detection above the $5\sigma$ level. Observations at other orbital phases
did not reveal a detectable signal from the source \citep{Aharonian2009, HESS2013, Romoli2015, Aharonian2005b}.

Due to the visibility constraints of ground-based telescopes, during each observation campaign it was
only possible to probe parts of the orbit around each periastron passage. In 2004, \psrls was observed
mostly after the periastron, in 2007 mostly before it and in 2011 only a short observation window
of five days around $t_{\rm p}+\SI{30}{\day}$ was available. A double-peak profile of the VHE light
curve in the orbital phase range encompassing the first and second disc crossings, with a local flux minimum close to $\sim t_{\rm p}$, results when combining data from all those years
(see e.g. the discussion in \citealp{Kerschhaggl2011}). The similarity of this light curve to the profiles derived from
radio and X-ray observations suggests a common particle population and/or emitting region.


Sufficient coverage of the orbital phase around the time of periastron passage at \vhes was
lacking up to now, preventing a definite assessment of the double-peak
profile or of any possible super-orbital variability of the source at these energies. Furthermore,
neither the periastron passage nor the periods in which HE flares occur have been covered deeply so
far. This situation changed after a deep \hess observation campaign was
conducted in 2014 under favourable observation conditions. Observations of \psrls in
2017 are reported here for the first time. For this campaign, the observation conditions were not optimal,
and the campaign was restricted to a short time period about 40 d before periastron.

In Sect. \ref{Sect:Observations} of this paper, the \hess observations
conducted in 2014 and 2017 are described. Furthermore, descriptions of the
telescopes and different \hess observation modes are given. In Sect.~\ref{Sect:Analysis}, details of the VHE analysis
procedures used in this paper are provided.
The results of the analyses are reported in Sect.~\ref{Sect:Results},
including a sky map obtained with the \hess-II array (see Sect.~\ref{Sect:Analysis}), as well as the spectral
and timing characterisations of the source for all available data. A dedicated analysis of the \fl data
during the source periastron passage in 2011, 2014 and 2017 has also been performed. This LAT analysis and the results obtained
for each event are summarised in Sect.~\ref{Sect:Fermi_analysis}. In Sect.~\ref{Sect:Discussion}, the
outcomes of the analyses are discussed in a multi-wavelength context. The findings from the analyses are compared with
 theoretical models addressing the phase-folded light-curve profiles and the spectral and timing properties
of \psrls in the gamma-ray domain. Conclusions and perspectives are briefly outlined in Sect.~\ref{Sect:Conclusions}.

\section{Observations of \psrls with \hess}
\label{Sect:Observations}
\subsection{The \hess array and observation modes}
\label{sec:hess_technical}

\hess is an array of five telescopes designed to detect the Cherenkov light produced during the development of air
showers that are initiated by highly-energetic particles as they interact with particles of air in the upper atmosphere. \hess is located in
the Khomas Highland in Namibia at an altitude of \SI{1800}{\meter} above sea level. In its first phase (\hess I),
the array consisted of four identical imaging atmospheric Cherenkov telescopes (CT1--4) with a mirror diameter of \SI{12}{\meter}, positioned on
the corners of a square with a side length of \SI{120}{\meter}.
The \hess I
array provides a field of view (FoV) with a diameter of about $\ang{5}$ and an energy threshold of about \SI{100}{\gev}.
A detailed description of the \hess I array can be found in \cite{Aharonian2006}.
%


In 2012 a new telescope (CT5), featuring a \SI{28}{\meter} parabolic dish, was added to the centre of the array, initiating the phase \hess II. The Cherenkov camera of CT5 covers a FoV of about \ang{3.2}. Due to the large reflector area, the energy threshold of CT5 can be as low as $\sim\SI{20}{\gev}$, for example in the case of pulsed signals \citep{velaPsr}.


In 2016, the cameras of the CT1--4 telescopes underwent an extensive upgrade aiming to reduce their readout dead
time and to improve the overall performance of the array \citep{Giavitto2017}. Data taken with these new cameras
in 2017 are, however, not included in the analyses presented here, such that data obtained with CT5 are considered
exclusively in this case.

The heterogeneous composition of the \hess-II array allows for various observation and analysis modes. In this paper,
data obtained with CT5 and a minimum of three of the CT1--4 telescopes are referred to as CT1--5 stereo-mode data. Data
analysed with a minimum of three of the smaller telescopes but not CT5 are referred to as CT1--4 stereo
mode, whereas CT5 mono indicates CT5 standalone analyses.


The \psrls VHE data reported in this paper were all taken in a mode in which the \hess telescopes are
pointed towards two symmetric positions offset from the source by \ang{0.5} along right ascension. In case of the 2017
data set, a third pointing position south of \psrls was used for \SI{30}{\percent} of the observations.

\subsection{\hess observations in 2014 and 2017}

The \hess observation campaign in 2014 benefited from favourable observing conditions of \psrls at orbital phases close
to its periastron passage. The source was observable both before, during and after the time of periastron,
in particular also at the orbital phase during which a HE gamma-ray flare was detected in 2011 and 2014. A rich data
set was obtained, including 141 (151) observation runs (typically \SI{28}{\minute}-blocks of observations), corresponding
to a total acceptance corrected observation time of \SI{62.2}{\hour} (\SI{63.4}{\hour}) in CT5 mono
(CT1--5 stereo) mode. The runs were taken at mean zenith angles in the range \ang{41} to \ang{47}, with an average value of \ang{42}.
Data were taken during six periods, each corresponding to one moon cycle, with interruptions in the monitoring of the source
every \SI{28}{\day} lasting about \SI{4}{\day}.

In 2017, \psrls was observable for a much shorter time period. Observations were conducted only in the time range $t_{\rm p} - \SI{42}{\day} \leq t \leq t_{\rm p} - \SI{37}{\day}$.
These observations were mainly intended to cover a phase period in which a relatively high flux level was observed
during the 2014 campaign. A data set of \SI{6}{\hour} was collected, taken at zenith angles between
\ang{52} to \ang{62} with an average value of \ang{57}.


\begin{table*}[t]
  \begin{center}
    \begin{tabular}{ll|ccccc}
      \toprule
      & & \textbf{2004} & \textbf{2007} & \textbf{2011} & \textbf{2014} & \textbf{2017} \\
      \midrule
      & Start Date & Feb 27 & Apr 09 & Jan 10 & Mar 07 & Aug 10 \\
      & End Date & Jun 15 & Aug 08 & Jan 16 & Jul 21 & Aug 20 \\
      \midrule
      \multirow{3}{*}{\textbf{CT5 Mono}} & $N_{\textrm{Runs}}$ & - & - & - & 141 & 12 \\
      & $t_L$ / [\si{\hour}] & - & - & - & 62.2 & 6 \\
      & $\bar{\Theta}$ / [\si{\degree}] & - & - & - & 41.8 & 57 \\
      \midrule
      \multirow{3}{*}{\textbf{CT1--5 Stereo}} & $N_{\textrm{Runs}}$ & - & - & - & 151 & - \\
      & $t_L$ / [\si{\hour}] & - & - & - & 63.4 & - \\
      & $\bar{\Theta}$ / [\si{\degree}] & - & - & - & 41.8 & - \\
      \midrule
      \multirow{3}{*}{\textbf{CT1--4 Stereo}} & $N_{\textrm{Runs}}$ & 138 & 213 & 11 & 163 & - \\
      & $t_L$ / [\si{\hour}] & 57.1 & 93.9 & 4.8 & 68.1 & - \\
      & $\bar{\Theta}$ / [\si{\degree}] & 42.5 & 45.1 & 47.6 & 41.9 & - \\
      \bottomrule
    \end{tabular}
    \caption{Summary of \hess observations from 2004 to 2017 used in this paper.
   $N_{\rm Runs}$ is the number of runs passing quality selection cuts. The acceptance corrected observation time is referred to as $t_L$,
    and $\bar{\Theta}$ indicates the mean observational zenith angle of observations.
       }
    \label{tab:psr_data_sets}
  \end{center}
\end{table*}

In Table \ref{tab:psr_data_sets}, the total number of runs, observation times and average zenith angles
of the 2014 and 2017 data sets are summarised. Values are provided for both monoscopic analyses and for analyses
using CT1--4.

\subsection{The 2004, 2007 and 2011 \hess data sets}

Observations of \psrls during its periastron passage in 2004, 2007 and 2011 have been reported in \cite{Aharonian2005b},
\cite{Aharonian2009} and \cite{HESS2013}, respectively. These data are re-analysed using up-to-date analysis
techniques, the same used for the analysis of the 2014 and 2017 data sets, in CT1--4 stereo mode. This approach allows
for a consistent study of cycle-to-cycle variability and for performing stacked analyses (see Sect.~\ref{subsection:lightcurves}) of all available \hess observations of the source in the aforementioned mode. A total observation time of
\SI{57.1}{\hour}, \SI{93.9}{\hour} and \SI{4.8}{\hour} for the 2004, 2007 and 2011 data sets, respectively, is used in this paper, with
corresponding zenith angle ranges of \ang{40} to \ang{51}, \ang{40} to \ang{60} and \ang{44} to \ang{52}. A summary of those values is given in Table \ref{tab:psr_data_sets}.
These data sets can differ slightly from the ones used in previous publications, as improved quality selection
criteria are applied. In particular, more restrictive values of the atmospheric
transparency coefficient and a minimum of three participating telescopes per run are required for the analysis reported here.

The phase bin coverage for these years is rather inhomogeneous. In 2004, observations took place mainly at
$t \gtrsim t_{\rm p}$, while in 2007, most of the data corresponds to $t \lesssim t_{\rm p}$. A minor overlap
between the two exists at $t_{\rm p}-\SI{10}{\day} \leq t \leq t_{\rm p}+\SI{15}{\day}$. The smaller 2011 data set was recorded around
$t_{\rm p} + \SI{25}{\day}$, with no phase-folded overlap with either the 2004 or the 2007 observations.

\section{\hess data analysis}
\label{Sect:Analysis}

Regardless of the analysis mode, data were analysed with two different analysis
pipelines, each using distinct calibration methods \citep{Aharonian2006} and independent, advanced gamma-ray
reconstruction techniques. The first reconstruction pipeline, used to obtain the results reported in this paper,
is based on the {\it{Image Pixel-wise fit for Atmospheric Cherenkov Telescopes}} method
(ImPACT, \citealp{Parsons2014}). This pipeline is based on a likelihood fitting of camera pixel
amplitudes to image templates generated by Monte Carlo simulations. The standard cut configuration was used.
The second pipeline, used as a cross-check analysis in this paper, is based on the
{\it{Model Analysis}} method \citep{Naurois2009}, in which the camera images are
compared with a semi-analytical model using a log-likelihood minimisation technique.
Both pipelines can be used to analyse data in CT5 mono mode, not taking into account information from
CT1--4 even if they participated in the data taking, in CT1--4 stereo mode, even if CT5 participated in
the data taking, or in CT1--5 stereo mode. The choice of an analysis mode is usually based on the
telescope participation, but also on the goal of the analysis.

Observation runs were selected for both the main and internal cross-check analyses,
based on independent run quality selection cuts \citep{Aharonian2006}. Only runs passing cuts in both pipelines were used in this analysis.

Background emission produced by hadrons, electrons and diffuse gamma-ray emission, for example from the
nearby Galactic plane, is calculated from source-free regions close to
the source under study. For morphological studies and the production of significance maps, the background
is estimated for each pixel from a ring around the pixel position (Ring Background method,
\citealp{Berge2007}). An adaptive algorithm is applied to optimise the size of the ring
to avoid artificial excesses.

The background for spectral analyses is derived from \textit{OFF} regions with a similar
offset with respect to the camera centre as the source \textit{ON} region (Reflected Background,
\citealp{Berge2007}). This ensures that a similar acceptance for background events in the source and
background control regions is obtained. Similar to the morphological analysis, regions with known sources are
excluded a priori in the background estimation for spectral analyses.

Systematic uncertainties for stereo analyses reported in this paper are based on the procedure
described in \cite{Aharonian2006}. The systematic uncertainty on the flux is estimated to be at a level of
$\Delta\phi_{\rm stereo}\approx 20\%$, whereas the uncertainty on the spectral slope is taken to be
$\Delta\Gamma_{\rm stereo}\approx 0.1$.  For the mono analysis, the studies of PKS 2155--304 and
PG 1553+113 \citep{HESS_AGN_Mono} are used as reference. Systematic uncertainties on the flux are similar
to the stereo case ($\Delta\phi_{\rm mono}\approx\SI{20}{\percent}$), whereas systematic uncertainties
on the photon index are estimated to range from about $0.17$ (for PKS 2155--304) up to $0.65$ (for PG 1553+113).
While the photon index of \psrls is more similar to the one of PKS 2155--304, the observation conditions are
more similar to those of PG 1553+113 (higher zenith angle). To be conservative, a value of
$\Delta\Gamma_{\rm mono} = 0.3$ is adopted for the results presented here.

\section{\hess analysis results}
\label{Sect:Results}

\subsection{Sky map and source statistics}
\label{subsection:maps}

\begin{figure*}[ht]
  \centering
  \includegraphics[width=0.6\textwidth]{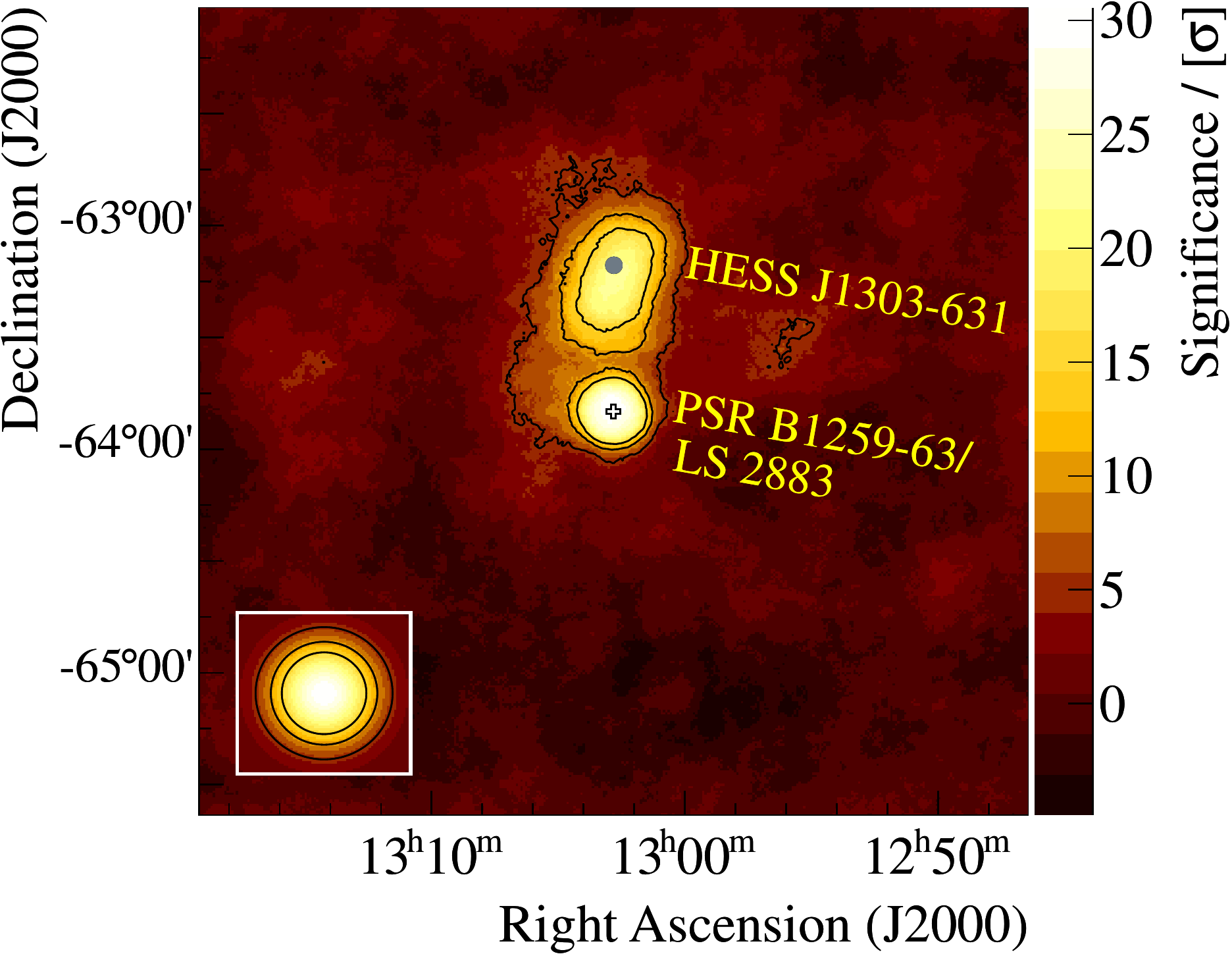}
  \caption{Significance map of the region around \psrls (white cross) obtained from the
           analysis of the 2014 data set in CT1--5 stereo mode. The pulsar wind nebula
           HESS~J1303--631 (best-fit ellipse centre \citep{j1303} indicated by a grey dot) is also detected
           to the north of \psrls. The bins in the map are correlated within a circle of radius \ang{0.14}.
           Significance contours are shown at the 5, 10, and 15$\sigma$ levels.  Some artefacts
           in the map are visible, occasionally exceeding the $5\sigma$ level, and are discussed
           in the text (see Sect.~\ref{subsection:maps}). In the inset, the point-spread function
           of the analysis is shown together with the respective contour lines.}
  \label{fig:maps}
\end{figure*}


The nominal position of \psrls is RA = $13^{\rm h} 02^{\rm m} 47^{\rm s}.65$, Dec = \ang{-63;50;08.6}).
A cut on $\theta^2 = 0.016$~deg$^2$, optimised for point-like source analyses, is used to evaluate
the statistics obtained on \psrls. A total of $15959$ gamma-ray candidate events are collected from the
\textit{ON} region around the source, whereas $75429$ such events are recorded from the \textit{OFF} background regions.
After correcting for the \textit{ON} to \textit{OFF} exposure ratio ($\alpha = 0.16$), a total of $3619$
excess events is obtained, yielding a detection of the source at a statistical significance of
36.5$\sigma$.

A significance map of the region around \psrls is shown in Fig.~\ref{fig:maps}.
This sky map is the first one produced for the source with the \hess-II array, and
corresponds to the analysis of the 2014 data set. The significance map corresponds to a total
acceptance corrected observation time of \SI{63.4}{\hour}, considering
events at energies above $E_{\rm th} = \SI{348}{\gev}$. The image and corresponding contours
were smoothed with a top-hat function on a scale similar to that of the PSF, which has a
\SI{68}{\percent} containment radius of \ang{0.1}. In addition to \psrls, the extended pulsar
wind nebula HESS J1303--631 \citep{j1303} is also apparent in the map, roughly one degree to the north from \psrls.
Residuals in the map are apparent and mostly related to background fluctuations.
Close to the two gamma-ray sources, towards the west in Fig.~\ref{fig:maps}, an additional
feature is visible at statistical significances of up to $5\sigma$
(significances here and thereafter are computed following \citealp{LiMa1983}).
This feature is caused by a block of sixteen adjacent pixels in the CT5 camera that was faulty during a large part of the
observations conducted in 2014. Furthermore, a gradient in the gains of the PMTs across the camera
of CT5 was present during the first part of the observation campaign, which contributes to the features
visible in the sky map. The influence of this feature on derived fluxes is estimated based on the variation of the
number of gamma-ray candidate events in each of the \textit{OFF} regions as a function of the
declination of the centre of the respective \textit{OFF} region. It is found that the effect on flux levels measured at the
position of \psrls is approximately \SI{1.5}{\percent}. This systematic effect is accounted for in the analysis of
the 2014 data, although it is noted that it is much lower than the commonly assumed level of
systematic uncertainties of \SI{20}{\percent} \citep{Aharonian2006} and thus negligible.


%

The data set recorded in 2017 is much smaller, as indicated in Table \ref{tab:psr_data_sets}. The analysis of this
data set results in a detection with a statistical significance at the level of $3.0\sigma$.

\subsection{Spectral analysis results}
\label{subsection:spectra}

\begin{figure*}[ht]
\begin{subfigure}{.5\textwidth}
  \centering
   \includegraphics[width=\textwidth]{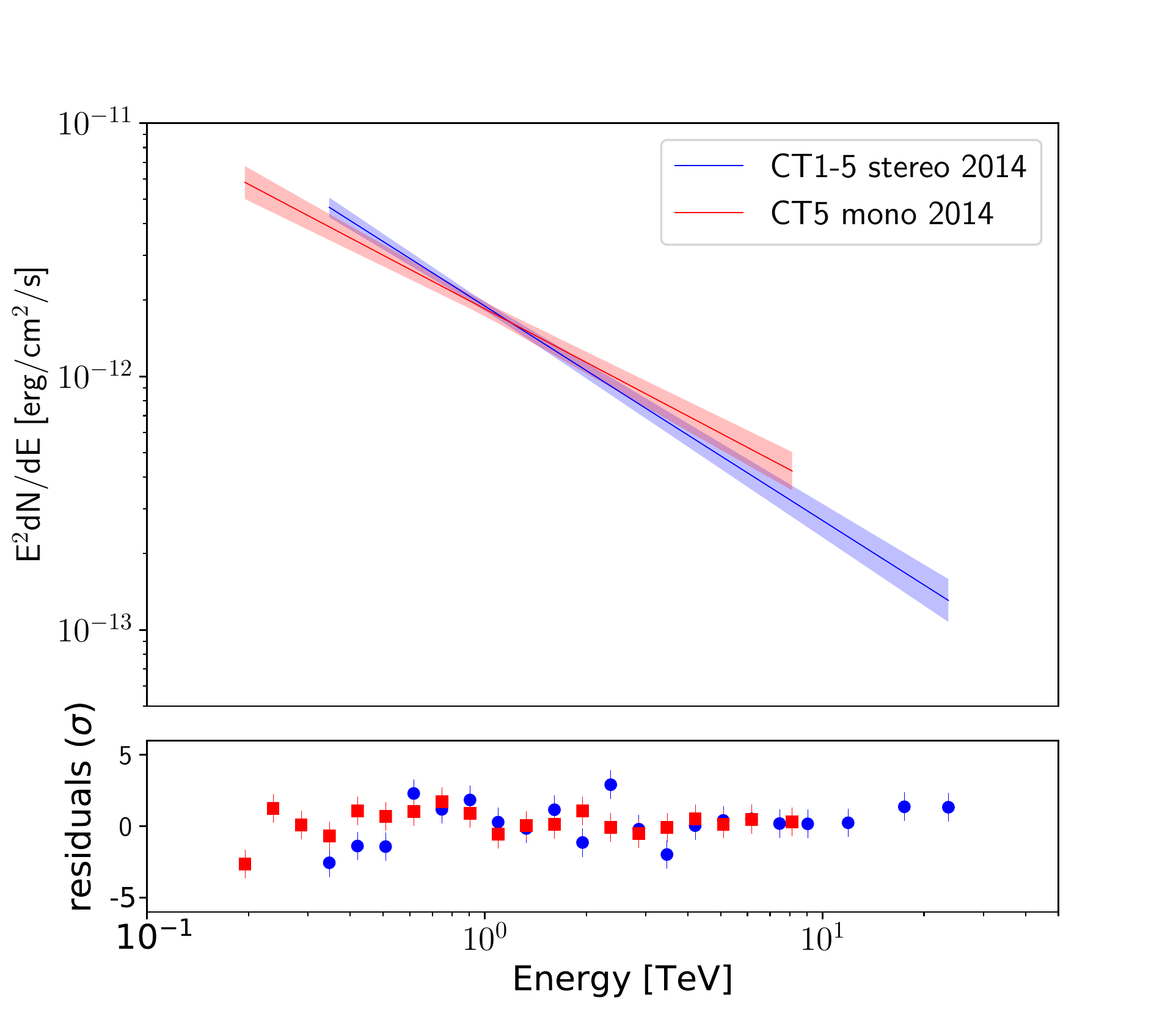}
\end{subfigure}%
\begin{subfigure}{.5\textwidth}
  \centering
  \includegraphics[width=\textwidth]{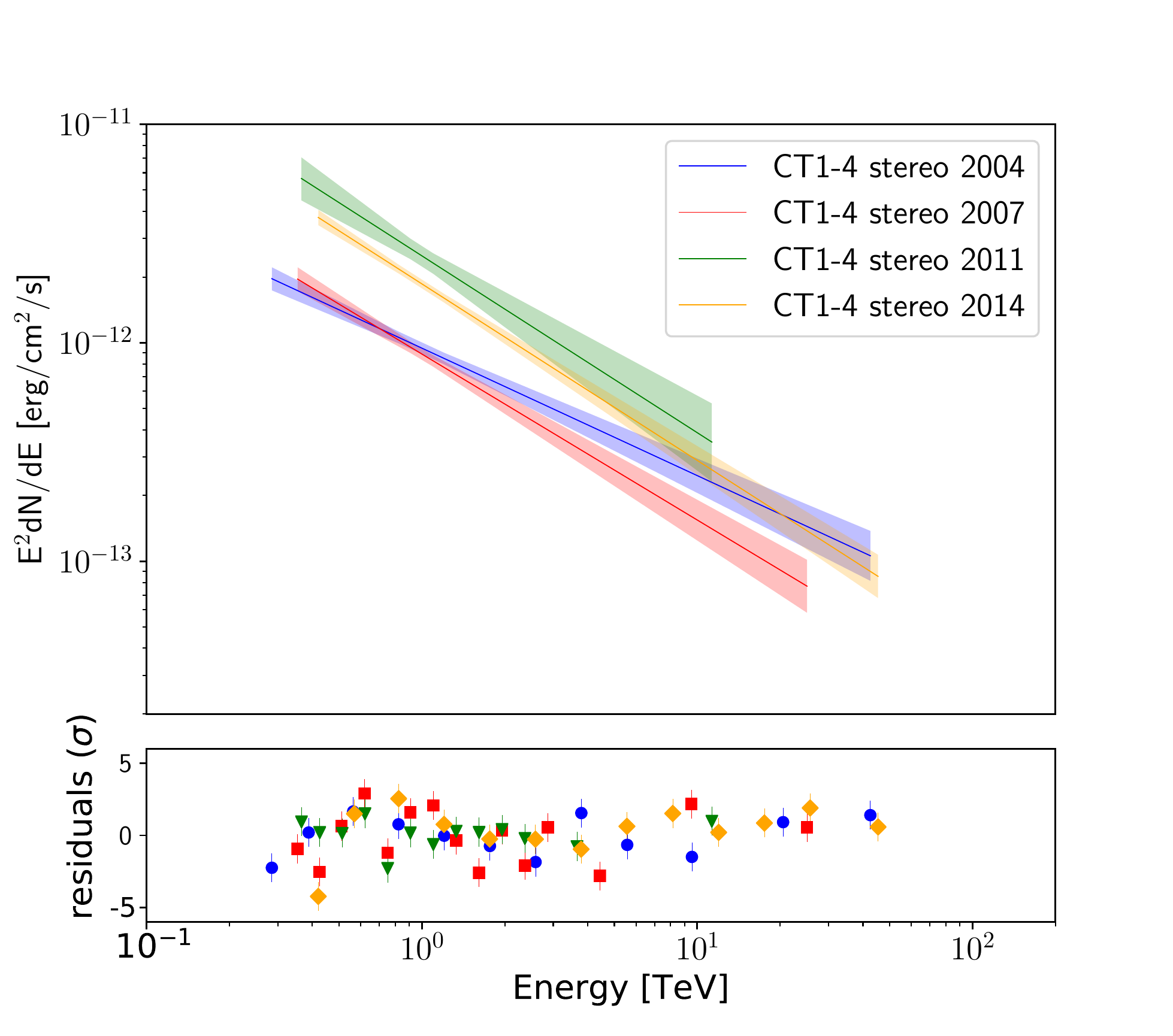}
\end{subfigure}
%
%
\caption{Spectral energy distribution of gamma rays from \psrls.
\textit{Left}: Results obtained from the observation campaign in 2014, with two different instrument configurations: CT5 mono and CT1--5 stereo, shown in red and blue, respectively.
\textit{Right}: Results from observation campaigns in 2004, 2007, 2011 and 2014 in CT1--4 stereo mode. Spectral energy bins are defined such that flux points have a statistical significance above $2\sigma$ in all cases. The highest-energy flux point of the 2014 spectrum is significant at the $1.5\sigma$ level only. Upper panels show the best power-law fit to the data, with coloured bands indicating the 1$\sigma$ statistical uncertainties on these fits. Lower panels provide the residuals of actual data to the best spectral fit. See main text for details.}

%
\label{fig:HESS_SED}
\end{figure*}

Spectra derived from the 2014 data set have been computed for the mono and CT1--5 stereo configurations. Figure \ref{fig:HESS_SED} (left panel) displays the spectral energy distributions for both configurations. In mono mode,
the spectral analysis covers the energy range $\sim \SI{0.18}{\tev}$--\SI{10}{\tev}, whereas the
stereo spectrum covers a range of \SI{0.32}{\tev}--\SI{26}{\tev}. Energy ranges for spectral analyses are defined
such that the energy reconstruction bias, determined from Monte Carlo simulations, is lower than
\SI{10}{\percent} of the energy, and in addition the effective area calculated for each data set
individually is required to exceed \SI{10}{\percent} of the maximum value.
The energy binning for the 2014 mono and stereo spectra shown in Fig.~\ref{fig:HESS_SED}
has been chosen such that every flux point has a statistical significance of at least $2 \sigma$.
In case of the data taken in 2017, no spectrum can be reconstructed due to the low number of gamma rays recorded.

Power-law models describe well both the mono and stereo spectra. Photon indices of $\Gamma_{\rm mono} = 2.70
\pm 0.05_{\rm stat} \pm 0.30_{\rm sys}$ and $\Gamma_{\rm stereo} = 2.88 \pm 0.05_{\rm stat}
\pm 0.10_{\rm sys}$ are obtained for the mono and stereo analyses, respectively.
Differential fluxes at \SI{1}{\tev} are found to be $\phi_{\rm mono}(\SI{1}{\tev}) = (1.85 \pm
0.11_{\rm stat} \pm 0.37_{\rm sys}$)$\times \SI{e-12}{\per\tev\per\second\per\centi\meter\squared}$ and
$\phi_{\rm stereo}(\SI{1}{\tev}) = (1.93 \pm 0.07_{\rm stat} \pm 0.38_{\rm sys}$)$\times \SI{e-12}{\per\tev\per\second\per\centi\meter\squared}$. The spectral results obtained in 2014 for both analysis configurations are compatible with
each other within uncertainties. No significant improvement, based on the comparison of likelihood values obtained from the
corresponding spectral fits, is attained when fitting the VHE data with more complex models.

Also the spectral energy distributions for the 2004, 2007 and 2011 periastron passages were computed. The results of the spectral analysis reported
here are compatible with previous reports for the corresponding years. Data from 2014, analysed in CT1--4 stereo mode, have also been included. The results are
displayed in the right panel of Fig.~\ref{fig:HESS_SED}, and fit details are given in
Table \ref{tab:hessI_spectra}. The distribution of photon indices obtained from these analyses are compatible
with a constant at the $2\sigma$ level. Hence it is concluded that there is no evidence
for super-orbital spectral variability. However, the flux normalisations obtained for the
different data sets display differences up to a factor of a few.
Such differences are attributed to different orbital phase coverage in each campaign (see Sect.~\ref{subsection:lightcurves}).

\begin{table*}[t]
  \begin{center}
    \begin{tabular}{c|ccc}
      \toprule
      Data Set & $\Gamma$ & $\phi(\SI{1}{\tev})$ / $\left[\SI{e-12}{\per\tev\per\second\per\centi\meter\squared}\right]$ & $E_{\textrm{C}}^{\SI{95}{\percent}}$ / [\si{\tev}] \\
      \midrule
      2004 & $2.64 \pm 0.06_{\rm stat} \pm 0.10_{\rm sys}$ & $0.97 \pm 0.05_{\rm stat} \pm 0.19_{\rm sys}$ & 22.4 \\
      2007 & $2.84 \pm 0.08_{\rm stat} \pm 0.10_{\rm sys}$ & $0.93 \pm 0.05_{\rm stat} \pm 0.19_{\rm sys}$ & 18.0 \\
      2011 & $2.7 \pm 0.1_{\rm stat} \pm 0.1_{\rm sys}$ & $2.4 \pm 0.3_{\rm stat} \pm 0.5_{\rm sys}$ & 1.5 \\
      2014 & $2.84 \pm 0.05_{\rm stat} \pm 0.10_{\rm sys}$ & $1.89 \pm 0.07_{\rm stat} \pm 0.38_{\rm sys}$ & 39.0 \\
      2014: CT1--5 Stereo & $2.88 \pm 0.05_{\rm stat} \pm 0.10_{\rm sys}$ & $1.93 \pm 0.07_{\rm stat} \pm 0.38_{\rm sys}$ & 10.1 \\
      \bottomrule
    \end{tabular}
    \caption{Spectral properties derived from the CT1--4 stereo (2004--2014) and  CT1--5 stereo analyses (2014). The photon index $\Gamma$, the differential flux $\phi$ at an energy of \SI{1}{\tev} and the derived lower limits on the cut-off energies $E_{\textrm{C}}^{\SI{95}{\percent}}$ are given. Limits are placed at the \SI{95}{\percent} confidence level.}
    \label{tab:hessI_spectra}
  \end{center}
\end{table*}

The CT1--4 spectra shown in Fig.~\ref{fig:HESS_SED} are well-described by simple
power-law spectral models. Due to the lack of cut-off features, lower limits on the energies at which
tentative exponential cut-offs could be located were derived for all
available spectra. The derivation is based on the simulation of \num{10000} spectra, simulated
under the assumption that the number of events in both the \textit{ON} and \textit{OFF} regions defined according to
the Reflected Background method follow a Poisson distribution. Systematic uncertainties
on the gamma-ray flux are taken into account in the simulation. Each of the resulting spectra
was fitted with a power-law function with an exponential cut-off. From the resulting set of energies
at which hypothetical cut-off values are placed, a lower limit on the cut-off energy at the
\SI{95}{\percent} confidence level was derived. The resulting values for each year can be found in Table \ref{tab:hessI_spectra}.
The most constraining value is found for the spectrum obtained from a CT1--4 analysis of the 2014 data set,
according to which a cut-off energy below \SI{39}{\tev} can be rejected at the \SI{95}{\percent}
confidence level. This finding strengthens the argument that particle acceleration is extremely efficient in \psrls.

\subsection{Light curve analysis results}
\label{subsection:lightcurves}

Phase-folded light curve profiles have been derived from the 2014 data set for CT5
mono and for CT1--5 stereo configurations. Three different integration time scales
have been considered: binning the data per night, week, and observing-cycle periods
(\SI{28}{\day}). Light curve points, including $1\sigma$ uncertainty intervals, are computed if the
respective flux estimate and the lower limit of its $1\sigma$ uncertainty interval are positive, otherwise a $68\%$ confidence level upper limit is calculated
assuming a power-law spectrum with an index derived from the spectral analysis described in Sect.~\ref{subsection:spectra}.

\begin{figure*}[t]
  \centering
  \includegraphics[width=\textwidth]{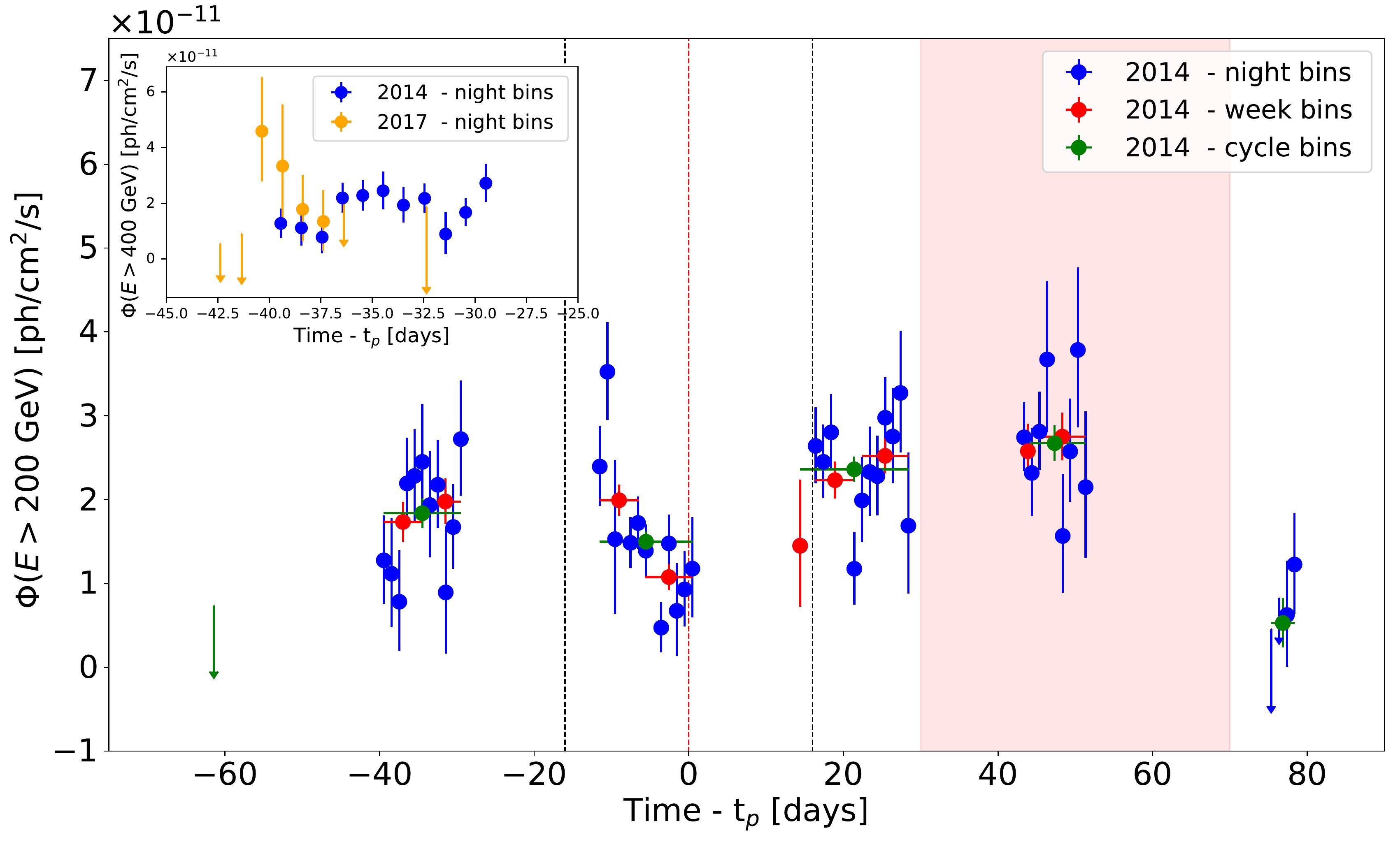}
  \caption{Light curve corresponding to the 2014 periastron passage recorded with CT5 mono configuration. Integral fluxes above \SI{200}{\gev} are displayed.
  Three different time binnings have been studied: days (blue), weeks (red) and observing periods (given by moon cycles of \SI{28}{\day},
  in green). Downward arrows are $68\%$ confidence level upper limits. Vertical dashed lines indicate the times of the disc crossings (grey)
  and the time of periastron $t_p$ (red). The red-shaded band indicates the period in which the HE gamma-ray flares have been reported in 2014.
  In the inset, the CT5 mono light curves obtained in 2014 and 2017 are shown above energies of \SI{400}{\gev}. The
  time range is limited to the part of the orbit sampled in 2017. In case of the 2017 light curve, a photon index of 2.7 is assumed.}
  \label{fig:HESS_LC_2014}
\end{figure*}

An energy threshold of \SI{200}{\gev} is employed for the CT5 mono analysis of the 2014 data set. The resulting flux points are shown in Fig.~\ref{fig:HESS_LC_2014}.

Light curves from the analysis of CT1--5 stereo data above an energy threshold
of \SI{400}{\gev} have also been studied. The shapes of these light curves are similar to the light curves shown
in Fig.~\ref{fig:HESS_LC_2014} and are therefore not reported in this paper.

For the 2017 data set, an energy threshold of the CT5 mono analysis of \SI{348}{\gev} is obtained.
Although the significance is less than the $5\sigma$ threshold for detection in this time period, a light curve has been computed assuming that the source emits VHE gamma rays,
as observed in previous observation campaigns at the same orbital phase. A photon index of $\Gamma = 2.7$ is assumed (see Sect.~\ref{subsection:spectra}).
The resulting light curve for integral fluxes above 400 GeV is shown for night-by-night integration time scales in the inset of Fig.~\ref{fig:HESS_LC_2014}, where the results for the 2014 data set in CT5 mono configuration for the same energy threshold are also included for comparison.

\begin{figure*}[t]
  \centering
  \includegraphics[width=\textwidth]{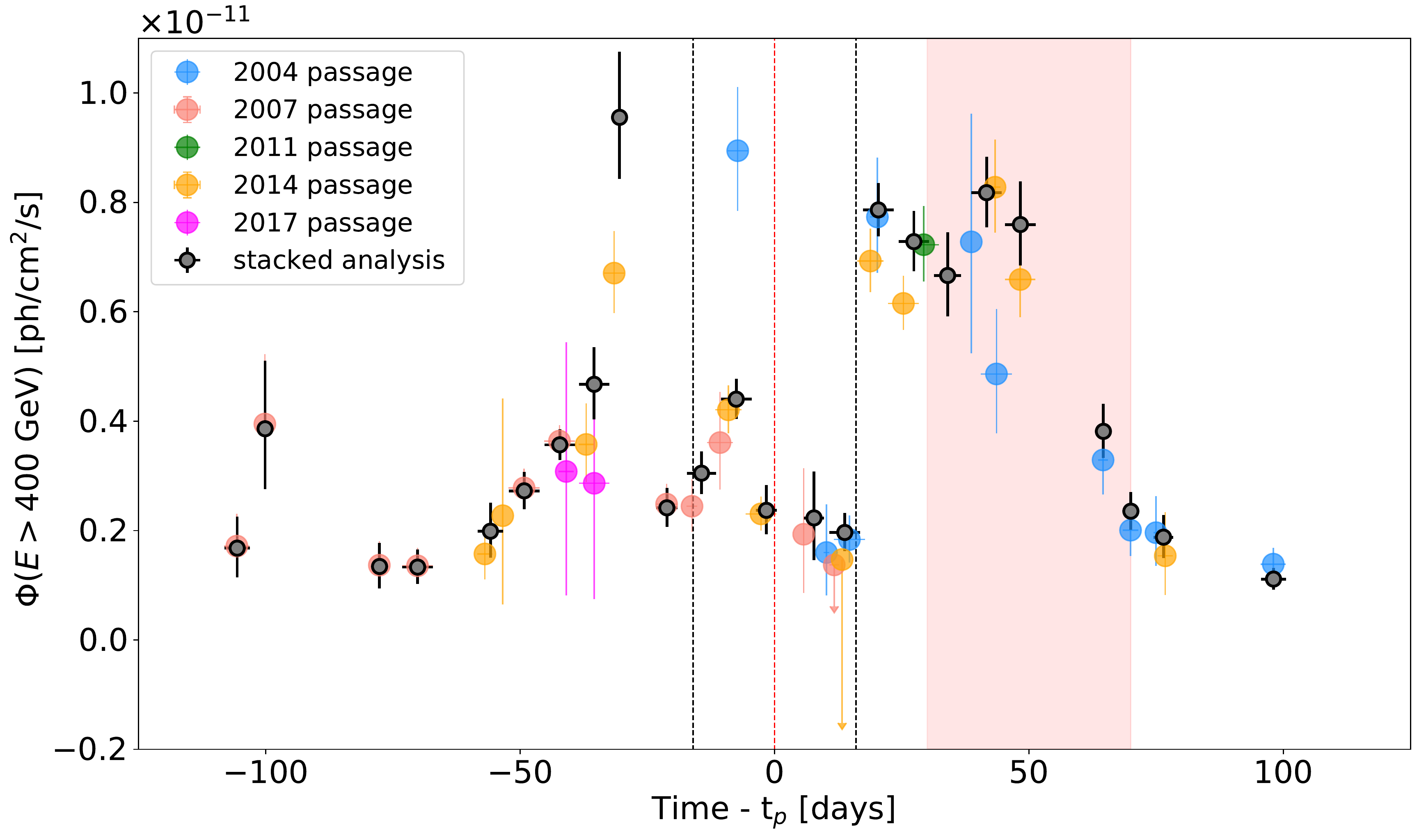}
  \caption{Weekly-binned light curves of the 2004, 2007, 2011, 2014 and 2017 periastron passages. Data from 2004 to 2011 have been reanalysed using the latest software available (the same as the one used for the 2014 analysis).
  A stacked light curve for all periastron passages in time bins
  corresponding to one week is also shown, with flux points derived assuming a photon index of 2.7 for all analyses.
  Downward arrows are $68\%$ confidence level upper limits. Vertical dashed lines and the red-shaded area are defined as in Fig.~\ref{fig:HESS_LC_2014}.}
  \label{fig:HESS-I_LC}
\end{figure*}

At the $3\sigma$ level, no evidence for short-term  (night-by-night) variability is
found when comparing adjacent bins within statistical uncertainties in the 2014 and
in the 2017 light curves. Light curves have also been produced using weekly integration time scales for the \hess-I
configuration including observations from 2004 to 2014 (see Fig.~\ref{fig:HESS-I_LC}).
These light curves have been computed by assuming photon indices obtained from annual averages.
Energy thresholds, defined here according to the aforementioned criterion on
the energy reconstruction bias only, are below \SI{400}{\gev} in all cases. Hence this energy
was adopted as a common threshold for the calculation of the light curves. For the 2017 data set,
results from the CT5 mono analysis are included in Fig.~\ref{fig:HESS-I_LC}.
Fluxes are compatible with the those observed at similar phases in previous years.

\subsection{Stacking analysis results}
\label{subsection:stacking}

Assuming that the processes responsible for the VHE emission from \psrls close to periastron repeat periodically,
a phase-folded stacking analysis including all \hess data sets available for \psrls has been performed.
The analysis is restricted to the CT1-4 stereo configuration for all observations, and a common energy threshold
$E_{\rm th} = \SI{400}{\gev}$ is applied. In Fig.~\ref{fig:HESS-I_LC}, weekly binned phase-folded light curves from
individual periastron passages (coloured flux points) are shown together with the all-year stacked light curve (grey flux points).
For the stacked light curve, a fixed photon index of $\Gamma = 2.7$ has been assumed.

Fitting the resulting light curve assuming a constant flux
yields an average value of $(2.8\pm0.4)\times \SI{e-12}{\photon\per\second\per\centi\meter\squared}$. The value $\chi^2/\textrm{Ndf} = 533/25$,
describing the fit probability given the number of degrees of freedom Ndf, implies that the fluxes are incompatible
with a constant at a level greater than $5\sigma$. Thus it is confirmed that \psrls is a variable source of VHE gamma rays.


Week-to-week variability is observed in different regions of the stacked light curve, in particular close to the time of the disc crossings.
The most significant flux variation is found at a level greater than $3\sigma$ at the second disc crossing between neighbouring bins
centred at \SI{11.5}{\day} and at \SI{17.8}{\day} after periastron. A flux increase by a factor of at least 2.9
is derived at the \SI{95}{\percent} confidence level.



\begin{figure*}
  \centering
  \includegraphics[width=\textwidth]{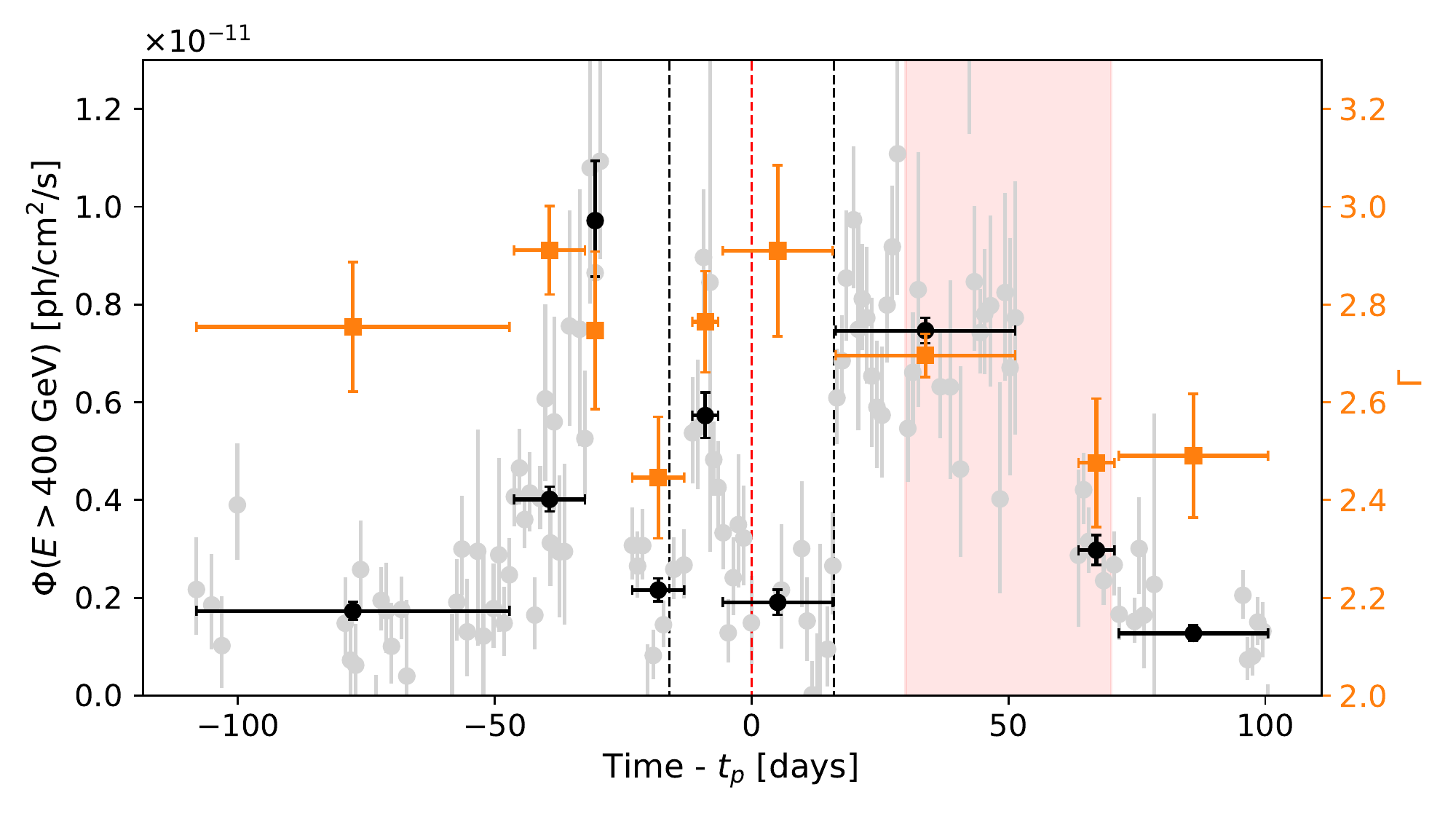}
  \caption{Integral flux above 400 GeV ($\Phi_{> \SI{400}{\gev}}$; black circles) and photon indices ($\Gamma$; orange squares) obtained from a stacked analysis of data corresponding to selected time intervals summarised in Table~\ref{table:stacked}. The light grey points correspond to the nightwise-binned phase-folded flux from which the Bayesian Blocks were computed. Vertical dashed lines and the red-shaded area are defined as in Fig.~\ref{fig:HESS_LC_2014}.
  }
  \label{fig:HESS_LC_bb_bins}
\end{figure*}

A stacked light curve has also been produced with a dedicated time binning that was derived from
a Bayesian Block analysis \citep{Scargle2013} on the phase-folded nightwise-binned light curve for energies above
\SI{400}{\gev}. The implementation in the {\tt astropy} Python package (version 2.0.12) was used, and
a false alarm probability of \SI{1}{\percent} was chosen. The aim is to identify flux states of the source that
could be linked with the physical processes taking place in the system.
The intervals found by the algorithm are illustrated in Fig.~\ref{fig:HESS_LC_bb_bins} and reported in
Table~\ref{table:stacked}. The identified intervals match the current knowledge of the orbital
behaviour of the system near the periastron and track the newly identified high state between \SI{-45}{\day} and \SI{-30}{\day}
from periastron. Distinct low flux states before the disc crossings, high flux states after the disc crossings
and a low state around periastron are identified.

The photon index in each interval is derived from a dedicated spectral analysis per bin. Spectral energy
ranges are defined according to the procedure described above. Photon indices for these spectra
are shown in Fig.~\ref{fig:HESS_LC_bb_bins}.

\begin{table*}[t]
  \centering
  \ra{1.3}
  \begin{tabular}{@{}crrrcrrrc@{}}
  \toprule
  Interval & Days from $t_{p}$ / [\si{\day}] & $t_L$ / $[\si{\hour}]$ && $\Gamma_{\rm VHE} $ & Signif. & $\Phi(E > \SI{400}{\gev})/\left[\SI[per-mode=fraction]{e-12}{\photon\per\centi\meter\squared\per\second}\right]$  \\
  \midrule
  Int. 1 & $\left[-108.085,-46.677\right]$ & 34.8 && $2.7\pm0.1$   & 11.0$\sigma$ & $1.7\pm0.2$ \\
  Int. 2 & $\left[-46.677, -31.944\right]$ & 31.2 && $2.91\pm0.09$ & 20.7$\sigma$ & $4.0\pm0.3$ \\
  Int. 3 & $\left[-31.944, -26.339\right]$ &  2.6 && $2.7\pm0.2$   & 12.1$\sigma$ & $9.7\pm1.2$ \\
  Int. 4 & $\left[-26.339, -12.334\right]$ & 29.4 && $2.4\pm0.1$   & 11.2 $\sigma$ & $2.2\pm0.2$ \\
  Int. 5 & $\left[-12.334,  -6.052\right]$ & 11.9 && $2.8\pm0.1$   & 16.5$\sigma$ & $5.7\pm0.5$ \\
  Int. 6 & $\left[-6.052,   16.208\right]$ & 29.0 && $2.9\pm0.2$   & 8.7$\sigma$ & $1.9\pm0.3$ \\
  Int. 7 & $\left[16.208,   57.470\right]$ & 42.6 && $2.70\pm0.04$ & 39.8$\sigma$ & $7.5\pm0.3$ \\
  Int. 8 & $\left[57.470,   71.057\right]$ & 12.6 && $2.5\pm0.1$   & 11.8$\sigma$ & $3.0\pm0.3$ \\
  Int. 9 & $\left[71.057,  100.526\right]$ & 29.3 && $2.5\pm0.1$   & 9.4$\sigma$ & $1.3\pm0.2$ \\
  \bottomrule
  \end{tabular}
    \caption{Phase-folded stacking analysis of all available data for dedicated time intervals identified with a Bayesian Block approach (see Sect.~\ref{subsection:stacking} for details).
 Columns indicate the time range with respect to periastron passage $t_{\rm p}$, the total
acceptance corrected observation time of the observations, the significance level of the stacked data set analysis, the photon index $\Gamma$,
and integral flux above \SI{1}{\tev} with statistical errors only (systematic uncertainties at the level  of $\sim$20\% and $\sim 0.1$ are to be considered for the flux and photon index values, respectively)}
  \label{table:stacked}
\end{table*}

\section{Fermi-LAT data analysis and results}
\label{Sect:Fermi_analysis}

Observations of \psrls have also been performed in HE gamma rays with the large area telescope (LAT)
onboard the {\it Fermi} satellite. As part of its full-sky coverage pointing strategy, the LAT monitored \psrls
during its last three periastron passages in 2010/11, 2014 and 2017 (see e.g.
\citealp{Abdo2011, Caliandro2015, Tam2018}). The source was clearly detected by the LAT, displaying
low to moderate fluxes around the disc crossings and at the periastron passage itself. In addition,
bright HE gamma-ray flaring episodes starting about 30--\SI{40}{\day} after periastron and lasting for roughly \SI{30}{\day}
have been detected after each of these three periastron passages. The nature of these flares is still unknown.


A re-analysis of the data of the 2010/11, 2014 and 2017 periastron passages using the
software provided by the \fl Collaboration (\textsl{Science Tools} \texttt{v10r0p5}, and the
instrument response functions (IRFs) \texttt{PASS8}) is reported. For each dataset, an interval of about 5 months was
chosen around the periastron passage. The time intervals are reported in Table~\ref{tab:fermi_res}.

Photon events in an energy range between \SI{100}{\mev} and \SI{500}{\gev} were extracted from a square region
with side lengths of \ang{28}, centred on the position of the \psrls system. The emission from this region of interest (RoI) was
modelled taking into account all the sources reported in the 3FGL catalogue \citep{Acero2015} within \ang{25}
of \psrls, also including the models for the Galactic and isotropic background emission (\texttt{gll\_iem\_V06}
and \texttt{ISO\_P8R2\_SOURCE\_V06\_v06}, respectively). A binned likelihood analysis was performed, with sources
outside the RoI having all their spectral parameters fixed to the 3FGL values. Sources included in the annulus
between \ang{5} and the edges of the RoI had only their spectral normalisation free to vary, while sources within
\ang{5} from \psrls had all their spectral parameters free. The fit was performed twice, taking into account
in the second fit only sources with a significance above 2$\sigma$. All the background sources were then fixed to the values
obtained through this procedure and a weekly light curve was then derived by performing a new likelihood fit
on weekly time intervals where only the flux normalisations of \psrls
and of sources flagged as variable in the 3FGL catalogue were left free to vary.
To better explore the emission in the pre-flare phase, a stacked analysis was also performed merging together all the data of the 3 passages.
To keep the computation of the light curve consistent, the data were stacked in a phase interval from \SI{66.5}{\day} before the periastron passage to \SI{24.5}{\day} after.
The result of the likelihood fits for each time interval is reported in Table~\ref{tab:fermi_res}.

\begin{table*}
\centering
\begin{tabular}{ccccc}
\toprule
Time Interval / [MJD] & Year & Days from $t_{\rm p}$ / [\si{\day}] & $\Phi (E>\SI{100}{\mev})$ / [\si{\photon\per\centi\meter\squared\per\second}] & $\Gamma$   \\
\midrule
55478.5 - 55646.5 & 2010/11&$\left[-66.5, 101.5\right]$ & $(3.6 \pm 0.3)\times10^{-7}$ & $2.98\pm0.09$ \\
56714.5 - 56882.5 & 2014 &$\left[-66.5, 101.5\right]$ & $(4.7 \pm 0.5)\times10^{-7}$ & $3.22\pm0.09$ \\
57951.5 - 58119.5 & 2017 &$\left[-66.5, 101.5\right]$ & $(4.6 \pm 0.3)\times10^{-7}$ & $2.98\pm0.07$ \\
stacked & all years &$\left[-66.5, 24.5\right]$ & $(2.0 \pm 0.3)\times10^{-7}$ & $3.0\pm0.2$ \\
\bottomrule
\end{tabular}
\caption{Likelihood fit results of the \fl data for the 2010/11, 2014 and 2017 periastron passages, together with the results of a stacked analysis for the pre-flare period. Columns indicate the integration time interval (in MJD), the phase interval is given as days from the periastron $t_p$, the integral flux is computed for energies above \SI{100}{\mev}, assuming a photon index $\Gamma$.}
\label{tab:fermi_res}
\end{table*}

\begin{figure*}[t]
  \centering
  \includegraphics[width=\textwidth]{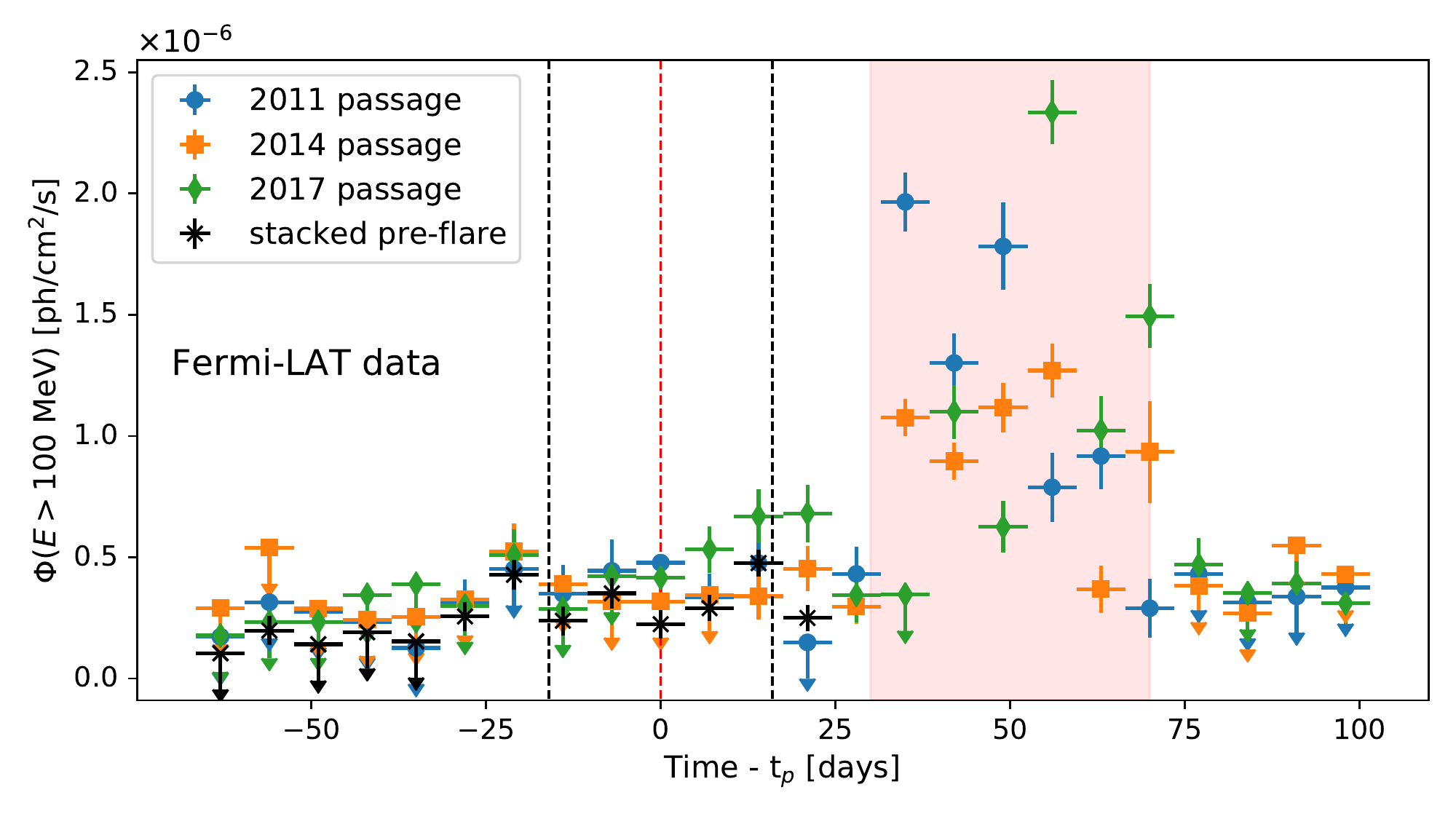}
  \caption{Light curves derived from \fl data for the periastron passages 2010/11 (blue circles), 2014 (orange squares), 2017
    (green diamonds), and the stacked analysis of all the data in the pre-flare phases (black crosses). Vertical dashed lines
    and the red-shaded area are defined as in Fig.~\ref{fig:HESS_LC_2014}.}
  \label{fig:LAT_LC_weekly}
\end{figure*}

The HE gamma-ray light curves with a weekly binning obtained from the analysis of \fl data are shown in
Fig.~\ref{fig:LAT_LC_weekly}. Flux points are displayed only if the significance of the source is above a TS
value of 10 (corresponding to a $\sim3\sigma$ detection), otherwise a 2$\sigma$ upper limit is placed.
The results of this analysis yields compatible results with what was reported previously \citep{Tam2018,Johnson2018,Chang2018}.
The most striking feature of these light curves are the bright HE flares starting about \SI{30}{\day} after periastron in 2010/11 and 2014, and $\sim\SI{40}{\day}$ after periastron in 2017. The flares last about a month in all cases with substructures on timescales from days to minutes \citep{Johnson2018}.

%

Using a stacked analysis of all the LAT observations of \psrls falling in the range [$t_p$ - \SI{100}{\day}, $t_p+\SI{25}{\day}$], we confirm the presence of significant emission during the two disc crossings and at $t_p$ \citep{Tam2018,Chang2018}.

\begin{figure*}[t]
  \centering
  \includegraphics[width=\textwidth]{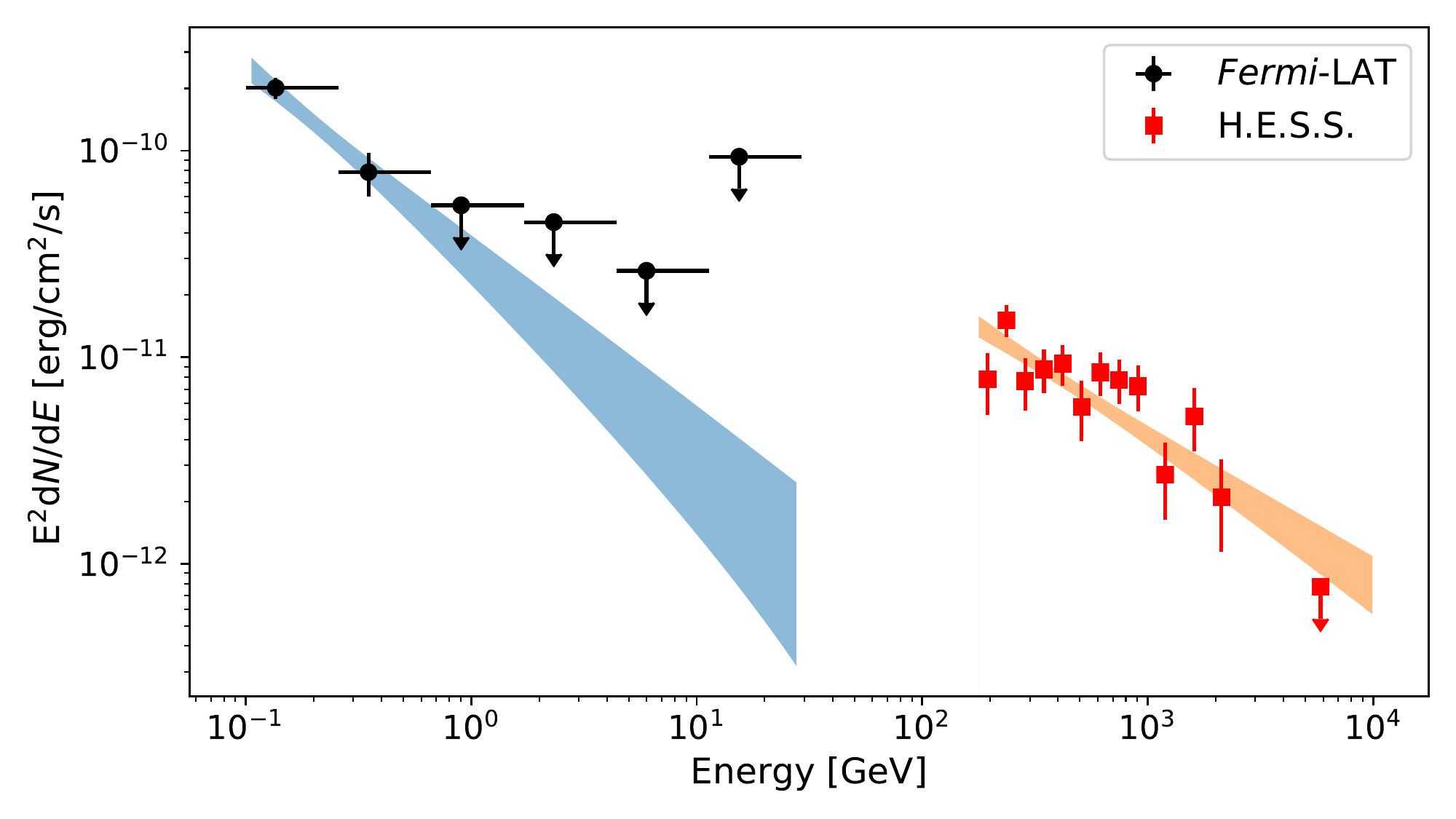}
  \caption{Spectral energy distribution of the emission from \psrls obtained with the \fl and \hess in CT5
    mono mode during the HE gamma-ray flare in 2014, based on observations taken between $t_p
    + \SI{44.2}{\day}$ and $t_p + \SI{52.2}{\day}$). Power laws are assumed for both data sets.}
  \label{fig:HESS_Fermi_SED}
\end{figure*}

Due to the simultaneity of the \hess observations with the ongoing GeV flare in 2014,
a dedicated spectral analysis was performed in the time interval from MJD 56825.2 to MJD 56833.2,
corresponding to the phase range [$t_p+\SI{44.2}{\day}$, $t_p+\SI{52.2}{\day}$] when the \hess data were taken.
The likelihood fit returned a flux above \SI{100}{\mev} of $\left(8.4\pm0.9\right) \times \SI{e-7}{\photon\per\second\per\centi\meter\squared}$
with a photon index $\Gamma = 2.93 \pm 0.15$. The SED obtained from this interval is shown in
Fig.~\ref{fig:HESS_Fermi_SED}. The spectral points were computed with a likelihood
fit in each energy band, where only the flux normalisation is left free to
vary while the photon index is fixed to the average value over the full energy range.

\section{Discussion}
\label{Sect:Discussion}

The long-term monitoring of \psrls allows its gamma-ray profile
to be robustly characterised, its HE and VHE spectral properties to be
constrained at different periastron passages and within distinct orbital
phases, and the (non-)correlation of the emission in these two energy bands
to be determined concretely.

It is noted that in this discussion only models employing leptons as primary particles are considered. The stellar
wind and the circumstellar disc provide reservoirs of target material with which relativistic protons could also
interact to produce gamma rays through $\pi^{0}$ decay. However, hadronic interpretations proposed for this
source \citep{Chernyakova2006, psr_hadronic_model} were not able to explain the light-curve profiles adequately
unless complex disc morphologies were invoked \citep{Aharonian2009}.
Furthermore, arguments based on the time scales of the observed variability favour leptonic
scenarios \citep{dubus2013}.

\subsection{Origin of short- and long-term VHE flux variability}
\label{sec:variability}

\psrls features distinct VHE maxima in its phase-folded light curve, contained within $\sim$ \SI{100}{\day} around the time
of periastron as shown in Fig.~\ref{fig:HESS-I_LC}. When all available data are combined, high emission states around
the times of the first and second disc crossings become apparent. This general trend is similar to and contemporaneous
with other non-thermal lower-energy profiles, as obtained from dedicated multi-wavelength analyses of the source (see
e.g. Fig.~5 in \citealp{Chernyakova2015}). A local VHE gamma-ray flux minimum is found at the time of periastron $t_p$.
Such a local double-peak profile centred approximately on $t_p$ and encompassing the emission at the times of the disc crossings
was already hinted at in previous \hess observations (see e.g.
\citealp{Aharonian2009}) and is confirmed here in particular with the data taken in 2014, which covered $t_p$ itself
for the first time.

In a scenario in which electrons are accelerated up to relativistic energies at the shock interface \citep{Rees1974,
Kennel1984} of the pulsar wind with the stellar wind emitted by LS 2883, VHE gamma-ray emission can be produced by
such electrons through inverse Compton (IC) scattering off the dense photon field provided by the bright companion star
(see e.g. \citealp{Tavani1997, Kirk1999}). In this framework, the relatively close distance between the compact object
and the companion star during orbital phases close to periastron is expected to cause enhanced gamma-ray fluxes from
the source. The presence of a double-peak profile encompassing the first and second disc crossings as well as the time
of periastron $t_p$, where a local minimum is found, has been successfully interpreted as a
signature of strong non-radiative losses in the system, which may be dominant all along the orbit (see e.g.
\citealp{Khangulyan2007, Kerschhaggl2011}). Adiabatic (or escape) losses displaying a peak at the times close to
periastron could reproduce the VHE gamma-ray profile. The origin of such a peak is still under debate.


Absorption of VHE photons can also be significant at these orbital phases. Recently, \cite{Sushch2017} evaluated such
effects while accounting for an improved description of the properties of the circumstellar disc (see e.g.
\citealp{vanSoelen2012}) and updated values for the companion star \citep{Negueruela2011}. These authors conclude
that a maximum of the $\gamma$--$\gamma$ attenuation should take place $\sim \SI{4}{\day}$ (\SI{2}{\day}) before
periastron for gamma-ray photons absorbed by photon fields of the star (circumstellar disc).
The nightwise-binned light curve shown in Fig.~\ref{fig:HESS_LC_2014} displays a minimum in the flux level a few days before periastron.
%
%
However, the statistical fluctuations of the time-binned flux points around this minimum make it difficult to further constrain its precise location.
In addition, the stacked light curves in Fig.~\ref{fig:HESS_LC_2014} and in Fig.~\ref{fig:HESS-I_LC} show that the emission level
remains relatively low for \SI{14}{\day} after periastron. At about $t_p + \SI{10}{\day}$, the optical depth should already
be about \SI{50}{\percent} of that at periastron, according to the predictions in \cite{Sushch2017}. Since
$\gamma$--$\gamma$ absorption alone cannot fully explain the dip of the VHE flux close to periastron, it becomes clear
that non-radiative losses and/or intrinsic changes in the emitter around periastron need to be taken into account as well.

With the extended data set analysed here, an asymmetry of the light curve peaks at VHEs becomes apparent (see
Fig.~\ref{fig:HESS-I_LC}). Comparing nightwise-binned fluxes observed at times symmetric to the time of periastron
yields the result that, on average, the fluxes measured during the second disc crossing phase are $2.0\pm0.1$ times
as high as the fluxes observed during the first disc crossing phase. The fine-tuning of
different emission or absorption mechanisms in order to obtain a symmetric profile at VHEs, for instance Doppler boosting
effects, $\gamma$--$\gamma$ absorption or a large emitter size, is no longer required (see
discussion in \cite{Kerschhaggl2011}). Such processes and their complex interplay may nevertheless still be present and
be partly responsible for the observed flux modulation.

The high fluxes observed at the times of the disc crossings rise within one day from local flux
minima just before the disc crossings. The decrease of flux levels is asymmetric, lasting few days in case of the peak
around the first disc crossing, and more than two weeks in case of the peak after periastron. \vhe fluxes measured
during the time of the HE flare connect smoothly with the flux observed after the second disc crossing.

Enhanced VHE emission on a relatively short time scale is also observed about \SI{15}{\day} before the time of the first disc
crossing ($\sim t_p - \SI{30}{\day}$). This is most evident from the observational campaign conducted with \hess in 2014.
This bright emission lasts for about five days and might therefore be unrelated to the emission produced at the disc
crossing itself. A similar enhancement is not observed at radio wavelengths, X-rays or HE gamma rays. The analysis of the data taken
in 2017 yields moderate fluxes compatible with the fluxes obtained from the 2014 data set at similar parts of the orbit.
Due to the uneven sampling of the orbit, the presence of the pronounced peak cannot be tested for periodicity.
Further observations of the source at VHEs are needed to clarify whether this is a repetitive event.


\subsection{spectral evolution and stacked analysis}
\label{sec:spectral_evolution}

No signature of spectral curvature has been found in any of the spectra shown in Sect.~\ref{subsection:spectra}, and
fitting with more complex spectral models has provided no significant improvement with respect to a power-law fit. The absence
of an energy cut-off implies a highly efficient acceleration process taking place in the system at orbital phases close
to the periastron passage. The severe radiation (and possibly non-radiative) energy losses do not imprint any feature on
the spectra measured by \hess A spectral break in the VHE gamma-ray spectra would be expected if absorption of VHE photons on the photon
field of either the companion star or its circumstellar disc were significant. This change in the spectral slope might however start already at lower energies,
such that the \hess data would correspond to the absorbed spectrum.

The observed VHE radiation from \psrls is likely produced through IC emission deep in the Klein-Nishina (KN) regime. For a black-body target photon field provided by the
companion star, with surface temperatures between $T_{\star} = \SI{27400}{\kelvin}$ and \SI{34000}{\kelvin} \citep{Negueruela2011}, the seed photons
have an energy of $h\nu_{\star} \approx 6.7$--\SI{8.3}{\ev}, respectively. The transition from the Thomson to the KN regime should take place at
corresponding electron Lorentz factors of $\Gamma_{\rm TH \rightarrow KN} \approx 1.9\times 10^4$ and $1.5 \times 10^4$, respectively. In the Thomson regime, the resulting
gamma-ray photon energy is typically $h\nu_{\gamma} \approx 4\, \Gamma_{\rm TH}^2 \, h\nu_{\star}$. For gamma-ray photons with energies of \SI{1}{\tev}, electrons
with Lorentz factors $\Gamma_{\rm TH} \approx 1.7$--$1.9 \times 10^5$ would be required. These Lorentz factors are much larger than $\Gamma_{\rm TH \rightarrow KN}$. If the emission from the
circumstellar disc is considered, assuming a Planck spectrum with a peak at about \SI{4.6}{\ev} (for a disc temperature of $T_{\textrm{disc}}\approx
\SI{19.000}{\kelvin}$), then $\Gamma_{\rm TH \rightarrow KN} \approx 2.7 \times 10^4$, whereas electrons with Lorentz factors $\Gamma_{\rm TH} \gtrsim 10^5$ would be required to produce \si{\tev} photons.
From this simple estimation, the existence of a spectral break following the transition from Thomson to KN regimes in the energy domain studied here can be ruled out.


The spectral results reported in Sect.~\ref{subsection:spectra} provide a 95\% confidence level constraint on the
minimum cut-off energy to \SI{39}{\tev} for the analysis of the 2014 data set. Being deep in the KN regime, with
$4\epsilon_{0}\Gamma \gg 1$ and $\epsilon_{0} = h\nu_{\star}/m_{e}c^2$ \citep{Moderski2005},
the energy of the emitting parent electron population reaches at least values similar to that of the IC
up-scattered photon, $\sim\SI{40}{\tev}$. This corresponds to a lower bound on the maximum electron Lorentz factor of
$\Gamma\gtrsim\num{8e7}$. In addition, the absence of a downturn of the VHE spectrum in this regime implies that no
transition from IC scattering to synchrotron emission as dominant energy loss processes occurs in the energy range
0.2--\SI{40}{\tev}.

The spectral profile plotted in Fig.~\ref{fig:HESS_LC_bb_bins} does not show a significant variability of the photon
index of \psrls within errors (statistical and systematic). This is in contrast to what is observed in the X-ray spectrum
\citep{Chernyakova2015}, where a hardening of the emission can be observed when the source approaches the periastron
passages, with softer emission observed at the disc crossings and at periastron itself. A hint for a softening of the spectral index at VHEs is
observed  right before periastron, but statistical and systematic errors prevent a firm
conclusion on this. It is worth noting that a hardening of the synchrotron X-ray emission would be expected in the
transition to the KN regime as a consequence of an excess at the high-energy end of the parent electron population. The
looser KN cross-section dependency on the electron energy would, however, compensate the IC contribution, such that
the effect would be limited in the VHE gamma-ray domain.

During the second disc crossing and during the HE flaring period, compatible photon indices of $\Gamma_{\rm VHE}
\sim 2.7$--2.8 are observed. Together with the smooth evolution of the VHE flux for $t \gtrsim t_p + \SI{20}{\day}$ (see
below), the evolution of the VHE spectral parameters around the time of the HE flaring period does not provide support
for a distinct emission component emerging at this orbital phase. It is worth noting, however, that in the HE regime the
photon index of the spectrum of \psrls is also unchanged for the whole periastron passage period and during the gamma-ray
flare. A spectral analysis of the stacked pre-flare data sets (see black data points in Fig.~\ref{fig:LAT_LC_weekly})
provides a photon index $\Gamma_{\rm HE} = 3.06 \pm 0.16$, well compatible with that obtained during the flaring periods (e.g.
$2.93 \pm  0.15$ in 2014; see also \citealp{Tam2018}). If the flare originated from a separate component showing up only after
$t \gtrsim t_p + \SI{30}{\day}$ in the HE regime, this cannot be distinguished based solely on the analysis of the photon index.

\subsection{The HE gamma-ray flare observed at VHEs}
\label{sec:he_flare_at_vhes}

The phase-folded VHE light curves reported in Figs.~\ref{fig:HESS_LC_2014} and \ref{fig:HESS-I_LC} (see also
Fig.~\ref{fig:HESS_LC_bb_bins}) display a pronounced high level of emission at $t \gtrsim  t_p + \SI{30}{\day}$, lasting for
at least three weeks. The flux transition from the orbital phases, in which the system is assumed to cross the circumstellar
disc of the companion star for the second time, is smooth. Furthermore the spectral properties remain unchanged (see
Fig.~\ref{fig:HESS_LC_bb_bins}). This suggests that the processes taking place in the system and responsible for the VHE
emission after this second disc crossing are related and seem to extend to longer timescales than previously expected. A
sudden increase by a factor of a few at these orbital phases, as seen in the HE gamma-ray band (see Fig.~\ref{fig:LAT_LC_weekly}),
is not observed at VHEs. Following the procedure in \cite{HESS2013}, a flare coefficient can be introduced and
constrained to quantify this statement. Based on the fluxes above \SI{200}{\gev} shown in Fig.~\ref{fig:HESS_LC_2014}, a flux
increase by a factor $\kappa > 2.1$ can be excluded at the \SI{95.4}{\percent} confidence level when comparing the monthly
bins just before and during the time of the GeV flare. At energies above \SI{1}{\tev}, the flare coefficient can be
constrained to $\kappa < 2.0$ based on the fluxes observed at the time of the second disc crossing and at the time of the HE
gamma-ray flare (Fig.~\ref{fig:HESS_LC_bb_bins}, Table ~\ref{table:stacked}). Systematic uncertainties have been taken into
account in the calculation of these limits. At HEs, instead, a flux increase from the last point before the flare to the first bin belonging to
the flare of a factor $\approx 4$ is observed.
Furthermore, the VHE emission shows only moderate variability at the time of the HE gamma-ray flare. This is in contrast to
the HE variability observed in \psrls in 2011, 2014 and 2017, when large flux differences down to variability time scales of
days and even hours \citep{Tam2018,Johnson2018} were observed.

From Fig.~\ref{fig:HESS_Fermi_SED} it also becomes apparent that distinct components are responsible for either the HE or the VHE
emission during the flare. The SED shows that an extrapolation of the power-law fit to the \fl data is incompatible with the emission
observed at VHEs by several orders of magnitude. Similar conclusions were derived from the analysis
of the limited VHE data set in 2011 \citep{HESS2013}. Contrary to the case of other gamma-ray binaries
\citep{Hadasch2012}, the sub-\si{\gev} emission is not accompanied by multi-\si{\gev} emission in \psrls.
Finally, it is noted that at other orbital phases, the HE flux of \psrls is too low for a similar spectral constraint.
%

The analysis of the \fl data reported in Sect.~\ref{Sect:Fermi_analysis} shows in addition that significant HE
gamma-ray emission is detected between the times when the compact object crosses the circumstellar disc.
It is considered unlikely that the mechanism responsible for this baseline emission is also responsible for the emission
during the HE gamma-ray flare. The true anomaly of the pulsar changes by \ang{180} in this time range
(see e.g. \cite{Aharonian2009} for a sketch), and the emission observed during the flaring period is thought to have
a strong geometric dependence.

Whether or not the HE and VHE gamma-rays (and X-ray and radio emission, if they are also correlated) at $t_p > \SI{30}{\day}$   
are produced by the same particle population, acceleration mechanism, or in the same physical regions, is still unclear.
There are no theoretical predictions of high VHE fluxes for such a remarkably long ($\gtrsim \SI{30}{\day}$) time scale after
the second disc crossing.

The HE and VHE gamma rays emitted between the disc crossings could have the same origin. The significant
differences of the flux profiles and spectral properties measured during the gamma-ray flare in
these two energy bands suggest that different processes are responsible for the emission in the
two domains at this part of the orbit.

\section{Conclusions}
\label{Sect:Conclusions}

The observations of the periastron passages of \psrls in 2014 and 2017 at VHEs with \hess together with a reanalysis of
data obtained in previous observations reveals a complex VHE gamma-ray light curve that remains repetitive over years.
A double-peak profile around $t_p$ encompassing the two disc crossing times emerges when all light curves are shown or stacked together, displaying a clear asymmetry between the pre- and post-periastron orbital phases. Additional features in the light curve are also apparent,
in particular those derived from the rich data set obtained in 2014. A high flux is observed $\sim$\SI{15}{\day} before the first
disc crossing. Subsequently, high flux levels are also observed during the HE flaring events. Contrary to the HE case,
however, there is a smooth continuation of the VHE flux from previous orbital phases. The \fl analysis reveals that
these HE flares exhibit a significant cycle-to-cycle variability. In particular for 2017, the onset of the flare is delayed
by $\sim \SI{10}{\day}$ with respect to 2011 and 2014. The analysis shows that HE gamma rays are produced also before
the flare, during the times of the disc crossing by the compact object and at $t_p$ itself. Therefore, any theoretical
interpretation correlating the emission at radio wavelengths, X-rays and VHE gamma-rays at orbital phases around $t_p$ should account as well for this emission.

There is no evidence for variability of the photon index at VHEs, therefore an hypothesis in which distinct
emission components dominate at different orbital phases around $t_p$ is not supported by the presented spectral
results. \psrls does not show spectral variability in our analysis at HE either.

A lower limit on the energy at which a hypothetical spectral cut-off could be found is placed at an energy of $\SI{39.0}{\tev}$.
Therefore, it is concluded that \psrls is a very efficient particle accelerator.

The next passage is expected in early 2021 ($t_p$ occurring on February 9). The source will
be observable in excellent conditions from the \hess site in Namibia.
Further observations of periastron passages of \psrls, in particular with the Cherenkov Telescope Array (CTA) \citep{cta,cta_science_book}, are expected to
substantially improve the VHE
characterisation of the source.


\section*{Acknowledgments}

The support of the Namibian authorities and of the University of Namibia in facilitating
the construction and operation of H.E.S.S. is gratefully acknowledged, as is the support
by the German Ministry for Education and Research (BMBF), the Max Planck Society, the
German Research Foundation (DFG), the Helmholtz Association, the Alexander von Humboldt Foundation,
the French Ministry of Higher Education, Research and Innovation, the Centre National de la
Recherche Scientifique (CNRS/IN2P3 and CNRS/INSU), the Commissariat a l'\'{E}nergie atomique
et aux \'{E}nergies alternatives (CEA), the U.K. Science and Technology Facilities Council (STFC),
the Knut and Alice Wallenberg Foundation, the National Science Centre, Poland grant no. 2016/22/M/ST9/00382,
the South African Department of Science and Technology and National Research Foundation, the
University of Namibia, the National Commission on Research, Science \& Technology of Namibia (NCRST),
the Austrian Federal Ministry of Education, Science and Research and the Austrian Science Fund (FWF),
the Australian Research Council (ARC), the Japan Society for the Promotion of Science and by the
University of Amsterdam. We appreciate the excellent work of the technical support staff in Berlin,
Zeuthen, Heidelberg, Palaiseau, Paris, Saclay, T\"{u}bingen and in Namibia in the construction and
operation of the equipment. This work benefitted from services provided by the H.E.S.S.
Virtual Organisation, supported by the national resource providers of the EGI Federation.


\bibliographystyle{aa} \bibliography{PSRB1259_AandA}

\end{document}